\documentclass[twocolumn, showpacs,amsmath,amssymb]{revtex4}
\usepackage{graphicx}
\usepackage{epstopdf}
\usepackage{dcolumn}
\usepackage{bm}
\usepackage{color}

\begin{document}

\preprint{APS/123-QED}

\title{Ultra-high Resolution Spectroscopy with atomic or molecular Dark Resonances:\\
Exact steady-state lineshapes and asymptotic profiles in the adiabatic pulsed regime}

\author{Thomas Zanon-Willette$^{1,2}$\footnote{E-mail address: thomas.zanon@upmc.fr \\}}
\address{$^{1}$UPMC Univ. Paris 06, UMR 7092, LPMAA, 4 place Jussieu, case 76, 75005 Paris,
France}
\address{$^{2}$CNRS, UMR 7092, LPMAA, 4 place Jussieu, case 76, 75005 Paris, France}
\author{Emeric de Clercq}
\affiliation{LNE-SYRTE, Observatoire de Paris, CNRS, UPMC, 61 avenue de l'Observatoire, 75014 Paris, France}
\author{Ennio Arimondo}
\affiliation{Dipartimento di Fisica "E. Fermi", Universit\`a di Pisa, Lgo. B. Pontecorvo 3, 56122 Pisa, Italy}
\affiliation{}

\date{\today}

\begin{abstract}
Exact and asymptotic lineshape expressions are derived from the
semi-classical density matrix representation describing a set of
closed three-level $\Lambda$ atomic or molecular states including
decoherences, relaxation rates and light-shifts. An accurate
analysis of the exact steady-state Dark Resonance profile describing the
Autler-Townes doublet, the Electromagnetically Induced
Transparency or Coherent Population Trapping resonance and the Fano-Feshbach lineshape,
leads to the linewidth expression of the two-photon
Raman transition and frequency-shifts associated to the clock transition. From an
adiabatic analysis of the dynamical Optical Bloch Equations in the weak field limit,
a pumping time required to efficiently trap a
large number of atoms into a coherent superposition of long-lived
states is established. For a highly asymmetrical configuration with different decay channels, a strong two-photon resonance based on a lower states population inversion is established when the driving continuous-wave laser fields  are greatly unbalanced.
When time separated resonant two-photon pulses are applied in the adiabatic pulsed regime for atomic or molecular clock engineering, where the first pulse is long enough to reach a coherent steady-state preparation and the second pulse is very short to avoid repumping into a new dark state, Dark Resonance fringes
mixing continuous-wave lineshape properties and coherent Ramsey oscillations are created. Those fringes allow interrogation schemes bypassing the power broadening
effect. Frequency-shifts affecting the central clock fringe computed from asymptotic profiles and related to Raman decoherence process, exhibit non-linear shapes with the three-level observable used for quantum measurement. We point out that different observables experience different shifts on the lower-state clock transition.
\end{abstract}


\maketitle

\section{Introduction}

\indent In the 1930s, molecular-beam magnetic resonance techniques achieved very high
precision allowing the observation of atomic/molecular
systems essentially in total isolation \cite{Kellogg:1946}. The
Rabi method revealed coupling interactions between internal energy states
and provided plenty of information not only on atomic and
molecular structure, but also on nuclear properties. In the 1950s, N.F. Ramsey realized a scheme with much higher resolution by increasing the
interaction time between the atom or molecule and the oscillating field \cite{Ramsey:1956}. Still today, this technique
provides the highest resolution in order to follow a dynamical evolution of
wave functions and probe its phase
accumulations.  Control and
elimination of systematic frequency shifts dephasing a wave
function oscillation at a natural Bohr frequency are
fundamental tasks to achieve precision measurement \cite{Cronin:2009}.\\
\indent An alternative tool to probe by the Rabi or Ramsey sequences,
for a dipole-forbidden transition, is to radiatively
mix the atomic or molecular states. As an example, without natural state
mixing from spin-orbit interaction, a long-lived Raman coherence between a ground state and a
long-lived (as in alkalis) or metastable (as in two-electron atoms) level, to be referred as clock states, can be established by a
two-photon process  via an upper excited level, thus forming a three-level $\Lambda$
system. The properties of such a system are strongly determined by an optical pumping mechanism leading to a
formation of a Dark Resonance associated with a trapping of the atomic population in a coherent superposition of states \cite{Hansch:1970,Brewer:1975,Whitley:1976,Arimondo:1976,Alzetta:1976,Gray:1978,Orriols:1979}.
Since such quantum superposition states are radiatively stable, the associated Raman coherence production leads to extremely narrow Dark Resonances allowing high-resolution frequency measurements \cite{Shah:2010}. Such coherences were explored for single trapped ions \cite{Siemers:1992,Janik:1985}, microwave clocks \cite{Vanier:2005}, microwave chips \cite{Treutlein:2004,Farkas:2010}, optical lattice clocks \cite{Santra:2005}, multi-photon excitations \cite{Hong:2005,Champenois:2007}, or nuclear clocks \cite{Peik:2002}. Similar coherent
superpositions are used in solid-state physics for quantum information \cite{Greilich:2006}, in super-conducting circuits \cite{Dutton:2006,Kelly:2010}, in a single impurity ion inserted into a crystal \cite{Santori:2006}, in quantum dots~\cite{Xu:2008} with protection against random nuclear spin interactions~\cite{Issler:2010}, and in opto-mechanical systems~\cite{Weis:2010,Safavi:2011}. They are also actively considered within the future challenge of realising nuclear systems for quantum optics  in the X-ray region~\cite{Coussement:2002,Burvenich:2006}. Dark Resonances, largely exploited in Quantum Optics, have been extended to the preparation of molecules in  ro-vibrational ground states \cite{Mark:2009}, and to coherent superposition of  atomic-molecular states in order to measure atomic scattering lengths and lifetimes of exotic molecular states \cite{Winkler:2005,Dumke:2005,Moal:2006,Moal:2007}.\\
\indent Three-level narrow resonances are associated to quantum interferences produced by amplitude
scattering into different channels and are strongly dependent on the configuration
intensities and detunings \cite{Lounis:1992,Stalgies:1998}. Such different lineshapes were
associated to quantum interferences as the Autler-Townes (AT) splitting of the resonance
\cite{Autler:1955}, the Fano-Feshbach (FF) canonical form \cite{Fano:1961}, the Dark Resonance (DR) lineshape \cite{Alzetta:1976} known
as Coherent Population Trapping (CPT) \cite{Whitley:1976,Arimondo:1996} or
Electromagnetically Induced Transparency (EIT) \cite{Harris:1990,Fleischhauer:2005}.\\
\indent Accurate calculation of the lineshape for a quantum superposition resonance
requires numerical integration of the Bloch's equations. The
literature reports on efforts to establish
approximate analytic equations applicable to each particular case and, in a few cases, exact but rather complex~\cite{Brewer:1975,Orriols:1979,Swain:1980,Swain:1982,Kelley:1994,Wynands:1999,Lee:2003,MacDonnell:2004}. The application of $\Lambda$ schemes to high accuracy atomic clocks, in microwave or optical domains, requires to determine precisely the physical processes affecting the resonance lineshape and the shifts of the clock frequency. For that purpose this work provides a careful analysis of the lineshape dependence
on different parameters characterizing the atomic or molecular
system under investigation.\\
\indent  The standard clock interrogation of a three-level system
involves continuous excitation of the two lower states while sweeping through
the Raman resonance. For that regime, starting  from the steady-state analytical solution
of three-level Optical Bloch Equations we derive the exact expression of the
resonance lineshape where the role of the relaxations and dephasing rates
determining the absorption profile is expressed with physical
meaning. Our detailed discussion of key lineshape parameters expands
previous analysis~\cite{Whitley:1976,Orriols:1979,Brewer:1975,Janik:1985,Lounis:1992,Stalgies:1998}.
We show that Autler-Townes, CPT, EIT and FF
lineshapes are associated to a universal two-photon resonance lineshape depending on system parameters~\cite{Zanon-Willette:2005}.
The analytical expressions for the frequency shift associated either to the FF extrema
or to the EIT resonance point out dependencies not obvious in a perturbation treatment. \\

\indent An alternative clock operation scheme is based on a Raman-Ramsey scheme with the application of time separated but
resonant two-photon pulses, experimentally introduced in the microwave domain \cite{Zanon:2005-PRL,Zanon:2005-IEEE} and extended to rubidium cold atoms~\cite{Chen:2010}. This operation was inspired by the Ezekiel's group work at MIT on a thermal beam of sodium atoms \cite{Thomas:1982}. While in the standard Ramsey approach, a coherent superposition of clock states in the bare atomic/molecular basis is dynamically produced by a $\pi/2$ pulse depending on pulse duration and laser power, the coherent superposition, in the three-level two-photon approach, is created by an optical pumping process long enough to reach a steady-state. This scheme overcomes the power broadening mechanism of the continuous wave resonance allowing to obtain high
contrasted signals in a saturation regime. This idea was extended in refs.~\cite{Zanon-Willette:2006,Yoon:2007,Yudin:2010} to the realization of EIT-Raman (and hyper-Raman) optical clocks with alkaline-earth-metal atoms. The time-separated and individually tailored laser pulses may be designed to create an atomic coherent superposition while eliminating
off-resonant ac Stark contributions from external levels modifying the optical clock resonance \cite{Zanon-Willette:2006}. For the regime of the first laser pulse long enough to produce an efficient coherent superposition, we present here a detailed analysis describing the dependence of the DR lineshape on the system parameters.\\
\indent Within the quantum clock framework, the determination of lineshapes and resonance shifts in
different experimental configurations remains an important issue, to be carefully investigated within the present work. An important result of the present
analytical and numerical analysis  for the resonance frequency-shift of a three-level quantum clock, is that different lineshapes versus the optical detunings are obtained depending on the experimentally detected population or coherence observable.\\
\indent The $\Lambda$ system and the Bloch's equations for an homogeneous medium are introduced
in Sec. II, where an adiabatic analysis of the time dependent equations determines the approximated time
scale required to produce an efficient atomic/molecular coherent superposition. Sec. III establishes an exact treatment of the excited state steady-state regime and derives  the key informations on
the Dark Resonance lineshape. In Sect. IV, we derive the steady-state profile of a two-photon resonance between clock states. In Sec. V, we focus our attention to the Raman coherence lineshape observed between clock states. Finally Sec. VI analyzes DR fringes produced with resonant two-photon pulses separated in time mixing steady-state properties and Ramsey oscillations. A detailed analysis of the  fringe properties is derived in the adiabatic regime where the first pulse establishes a steady-state solution and the probe pulse duration vanishes. Instead only dynamical properties of these phase-shifts were demonstrated in refs \cite{Hemmer:1989,Shahriar:1997,Zanon-Willette:2006,Yoon:2007}. In the Appendix A, we rewrite lineshape population solutions in terms of generalized multi-photon transition rates enhancing one and two photon transition rates in the three-level system. We finally derive in Appendix B a first order analytical expression of the central fringe Raman frequency-shift associated to the pulsed Dark Resonance lineshape.

\section{Three-level Optical Bloch equations}

\begin{figure}[!t]
\centering
\resizebox{9.0cm}{!}{\includegraphics[angle=-90]{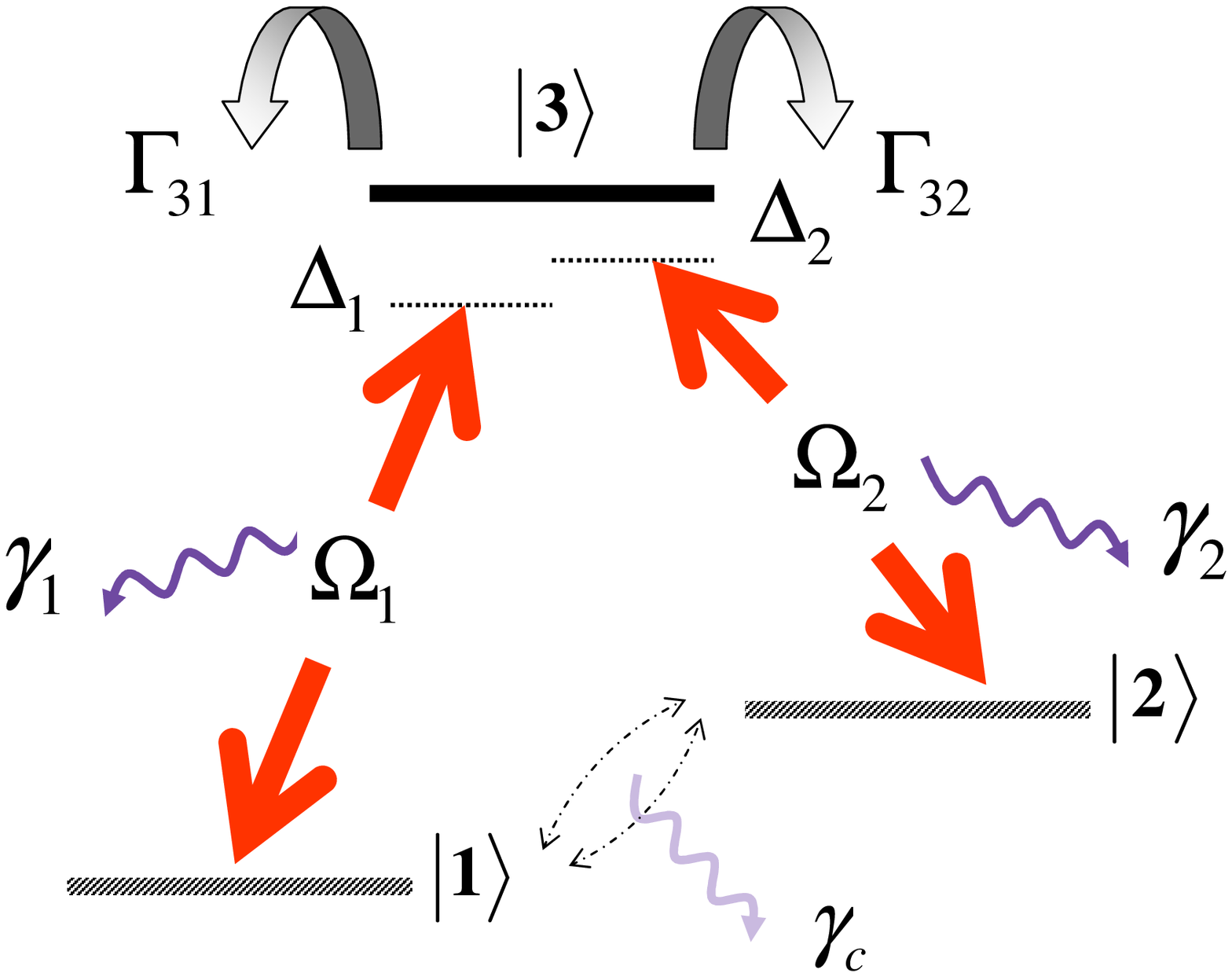}}
\caption{(Color online) Closed three-level $\Lambda$
configuration in the density matrix representation including relaxation rates
$\Gamma_{31},\Gamma_{32}$, and decoherences
$\gamma_{1},\gamma_{2},\gamma_{c}$. Optical detunings are
$\Delta_{1}$, $\Delta_{2}$. The parameter $\delta_{r}=\Delta_{1}-\Delta_{2}$
defines the Raman resonance condition. $\Omega_1$ and $\Omega_2$ define the couplings with the applied
laser fields. $|1\rangle$ and $|2\rangle$ are the clock states while $|3\rangle$ is the excited
state.}
\label{scheme-lambda}
\end{figure}
The Doppler-free three-level system presented in Fig.~\ref{scheme-lambda} is described by the density
matrix \mbox{$\rho_{ij}$ ($i,j=1,2,3$)} obeying the Liouville equation
\begin{equation}
\frac{d}{dt}\rho_{ij}=\frac{1}{i\hbar}\sum_{k}\left(
H_{i,k}\rho_{k,j}-\rho_{i,k}H_{k,j}\right)+\mathcal{R}\rho_{ij}.
\label{pilot-equation}
\end{equation}
The coupling of the three atomic or molecular states to two
coherent radiation fields, see Fig.~\ref{scheme-lambda}, is described within the
rotating wave approximation (RWA) by the following Hamiltonian:
\begin{equation}
H=\hbar\begin{pmatrix}
\Delta_{1}&0&\Omega_{1}\\
0&\Delta_{2}&\Omega_{2}\\
\Omega_{1}&\Omega_{2}&0\\
\end{pmatrix},\label{hamiltonian}
\end{equation}
where $\Delta_{1},\Delta_{2}$ are the detunings of the two fields.
Depending on the transition the Rabi frequencies $\Omega_1$ and $\Omega_2$ driving the system are
determined by the product either between electric dipole and electric field amplitude
or between magnetic dipole and magnetic field amplitude. It is worth noticing that
Rabi frequencies defined here are half of the definition of ref.~\cite{Cohen-Tannoudji:1998}. The
matrice $\mathcal{R}\rho_{ij}$ taking into account
relaxation and decoherence phenomena, is written
\begin{equation}
\mathcal{R}\rho_{ij}=
\begin{pmatrix}
\Gamma_{31}\rho_{33}&-\gamma_{c}\rho_{12}&-\gamma_{1}\rho_{13}\\
-\gamma_{c}\rho_{21}&\Gamma_{32}\rho_{33}&-\gamma_{2}\rho_{23}\\
-\gamma_{1}\rho_{31}&-\gamma_{2}\rho_{32}&-\Gamma\rho_{33}\\
\end{pmatrix}.\label{relaxation-matrix}
\end{equation}
The total spontaneous emission
rate $\Gamma$ is composed by the $\Gamma_{31},\Gamma_{32}$ rates (with
$\Gamma=\Gamma_{31}+\Gamma_{32}$) describing  either alkaline ($\Gamma_{31}\approx\Gamma_{32}$) or
akaline-earth ($\Gamma_{31}\neq\Gamma_{32}$) three-level
decay configuration. Optical coherences are relaxed with terms
$\gamma_{1},\gamma_{2}$ where
$\gamma_{1}+\gamma_{2}=\gamma$.
In a pure radiative process
\cite{Cohen-Tannoudji:1998}, optical decoherences are related to
spontaneous emission rates by the relation
$\gamma_{1}=\gamma_{2}=(\Gamma_{31}+\Gamma_{32})/2$. The $\rho_{12}$ decoherence
is described by the $\gamma_{c}$ dephasing term.
The optical Bloch equations describe the temporal evolution of the density matrix elements in the RWA as \cite{Zanon-Willette:2005}:
\begin{eqnarray}
\left\lbrace
\begin{split}
\frac{d\rho_{11}}{dt}=&-2\Omega_{1}Im\{\rho_{13}\}+\Gamma_{31}\rho_{33},\\
\frac{d\rho_{22}}{dt}=&-2\Omega_{2}Im\{\rho_{23}\}+\Gamma_{32}\rho_{33},\\
\frac{d\rho_{33}}{dt}=&~2\Omega_{1}Im\{\rho_{13}\}+2\Omega_{2}Im\{\rho_{23}\}-\Gamma\rho_{33},\\
\frac{d\rho_{13}}{dt}=&-[\gamma_{1}+i\Delta_{1}]\rho_{13}+i\Omega_{2}\rho_{12}-i\Omega_{1}(\rho_{33}-\rho_{11}),\\
\frac{d\rho_{23}}{dt}=&-[\gamma_{2}+i\Delta_{2}]\rho_{23}+i\Omega_{1}\rho_{21}-i\Omega_{2}(\rho_{33}-\rho_{22}),\\
\frac{d\rho_{12}}{dt}=&-[\gamma_{c}+i(\Delta_{1}-\Delta_{2})]\rho_{12}+i\Omega_{2}\rho_{13}-i\Omega_{1}\rho_{32},\\
\end{split}\right.
\label{set-Bloch}
\end{eqnarray}
with $\rho_{ji}=\rho_{ij}^{*}$. The population conservation of the closed system is given
by $\rho_{11}+\rho_{22}+\rho_{33}=1$. The optical detunings will be related to the Raman detuning $\delta_{r}$ by $\Delta_{1}=\Delta_{0}$ and
$\Delta_{2}=\Delta_{0}-\delta_{r}$. $\Delta_0$ represents the common optical detuning for a configuration where one laser is fixed while the other is frequency scanned. Notice that within the approach of deriving the $\Delta_1, \Delta_2$ laser modes from
a single source by modulation at frequency $\delta_r$, with  $\Delta_{1}=\Delta_{0}\pm\delta_{r}/2$ and $\Delta_{2}=\Delta_{0}\mp\delta_{r}/2$, the light-shift derivation should be modified.\\
\indent Eqs.~(\ref{set-Bloch}) describe the transient dynamics and the steady state of populations and quantum coherences.
A complete state mixing is reached when all atoms or molecules have been
pumped efficiently into the dark state, a coherent superposition of the lower states. Thus, a
pumping time is required to achieve an optimal atomic fraction trapped into the
coherent state superposition. We derive such time scale evolution from
Eq.~(\ref{set-Bloch}) by an adiabatic elimination of the time
derivative, $d\rho_{33}(t)/dt=d\rho_{13}(t)/dt=d\rho_{23}(t)/dt\equiv0$ for pulse durations greater than
$\Gamma_{31}^{-1},\Gamma_{32}^{-1}$, as the
population $\rho_{33}$ and optical coherences $\rho_{23}$ and
$\rho_{13}$ evolve more quickly than the populations $\rho_{11}$,
$\rho_{22}$ and the Raman coherence $\rho_{12}$. \\
\indent We investigate the dynamics of the three-level $\Lambda$ systems using various combination of long and short two-photon pulses separated in time.
A straightforward temporal analysis of the
resulting adiabatic set, similar to NMR equations
\cite{Jaynes:1955,Torrey:1949}, exhibits two damping times:
$\tau_{osc}$, determining the phase memory of
the Raman coherence precession (equivalent to a transversal or
spin-spin relaxation rate), and $\tau_{p}$ which determines
the typical population transfer into the dark state superposition (similar to a longitudinal or
spin-lattice relaxation rate) \cite{Abragam-Schoemaker}. At low optical saturation $\Omega_1,\Omega_2 \ll \Gamma, \gamma$, we have
\begin{eqnarray}
\begin{split}
\tau_{osc}(\Delta_0)&\sim\left(\gamma_{c}+\frac{\Omega_{1}^{2}}{\widetilde{\gamma_2}}+\frac{\Omega_{2}^{2}}{\widetilde{\gamma_1}}\right)^{-1},\\
\tau_{\rm p}(\Delta_0)&\sim\left(\frac{\Omega_{1}^{2}}{\widetilde{\gamma_1}}+\frac{\Omega_{2}^{2}}{\widetilde{\gamma_2}}\right)^{-1}\left[1+\widetilde{\Upsilon}
\frac{\Omega_{2}^{2}/\widetilde{\gamma_2}-\Omega_{1}^{2}/\widetilde{\gamma_1}}
{\Omega_{2}^{2}/\widetilde{\gamma_2}+\Omega_{1}^{2}/\widetilde{\gamma_1}}\right]^{-1},
\end{split}
\label{time-population}
\end{eqnarray}
where we have introduced the following generalized relaxation rates:
\begin{equation}\label{gammatilde}
\widetilde{\gamma_i}=\frac{\Delta_{i}^{2}+\gamma_{i}^{2}}{\gamma_{i}},i=1,2.
\end{equation}
The $\widetilde{\Upsilon}$ generalized branching ratio difference is
\begin{eqnarray}
\widetilde{\Upsilon}=\frac{\frac{3}{\Gamma}(\Omega_{1}^{2}/\widetilde{\gamma_1}
-\Omega_{2}^{2}/\widetilde{\gamma_2})+\Upsilon}
{\frac{3}{\Gamma}(\Omega_{1}^{2}/\widetilde{\gamma_1}
+\Omega_{2}^{2}/\widetilde{\gamma_2})
+1},\label{branching-ratio}
\end{eqnarray}
with the $\Upsilon$ normalized branching ratio given by
$\Upsilon=(\Gamma_{31}-\Gamma_{32})/(\Gamma_{31}+\Gamma_{32})$.
The $\tau_{\rm osc}$ and $\tau_{\rm p}$ timescales play a key role on the population transfer between atomic or molecular states \cite{Zanon-Willette:2006}.
Indeed, optical coherences are efficiently generated only when the $\Omega_1$ and $\Omega_2$ Rabi frequencies
are applied  for a time  $\tau$ exceeding $\Gamma_{31}^{-1},\Gamma_{32}^{-1}$. A
short pulse duration having $\Gamma_{31}^{-1},\Gamma_{32}^{-1}<\tau<\tau_{p}$ will be Fourier limited and will lead to a weak contrast
resonance profile, whereas a long pulse with
$\tau\gg\tau_{p}\gg\Gamma_{31}^{-1},\Gamma_{32}^{-1}$ will
eliminate all time dependencies in lineshape and frequency
shifts. This regime is latter examined within the following section.

\section{The steady-state lineshape of $\rho_{33}$}

\subsection{The Dark Resonance}

\begin{figure}[!!t]
\centering
\hfill\resizebox{8.5cm}{!}{\includegraphics[angle=0]{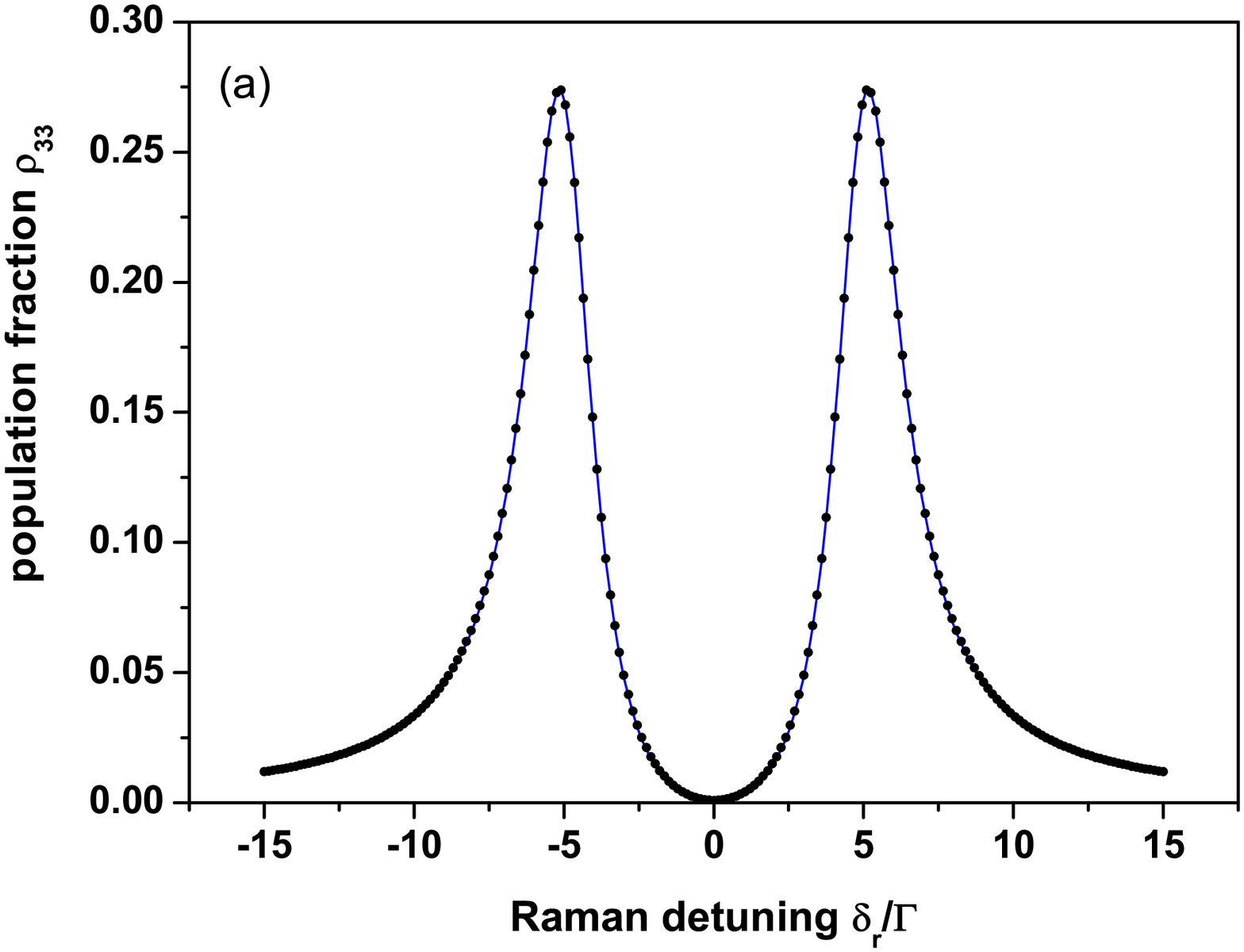}}
\hfill\resizebox{8.5cm}{!}{\includegraphics[angle=0]{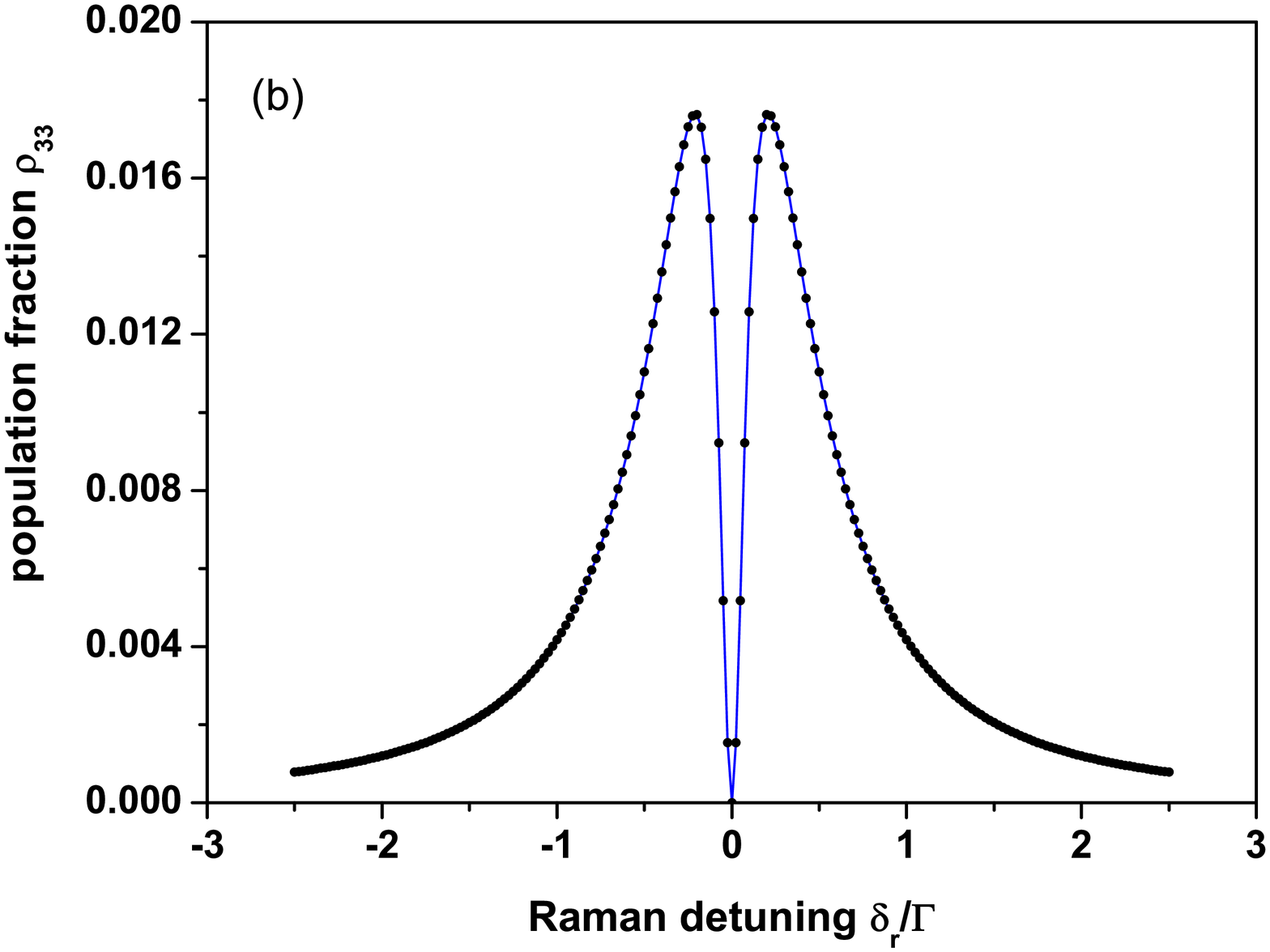}}
\hfill\resizebox{8.5cm}{!}{\includegraphics[angle=0]{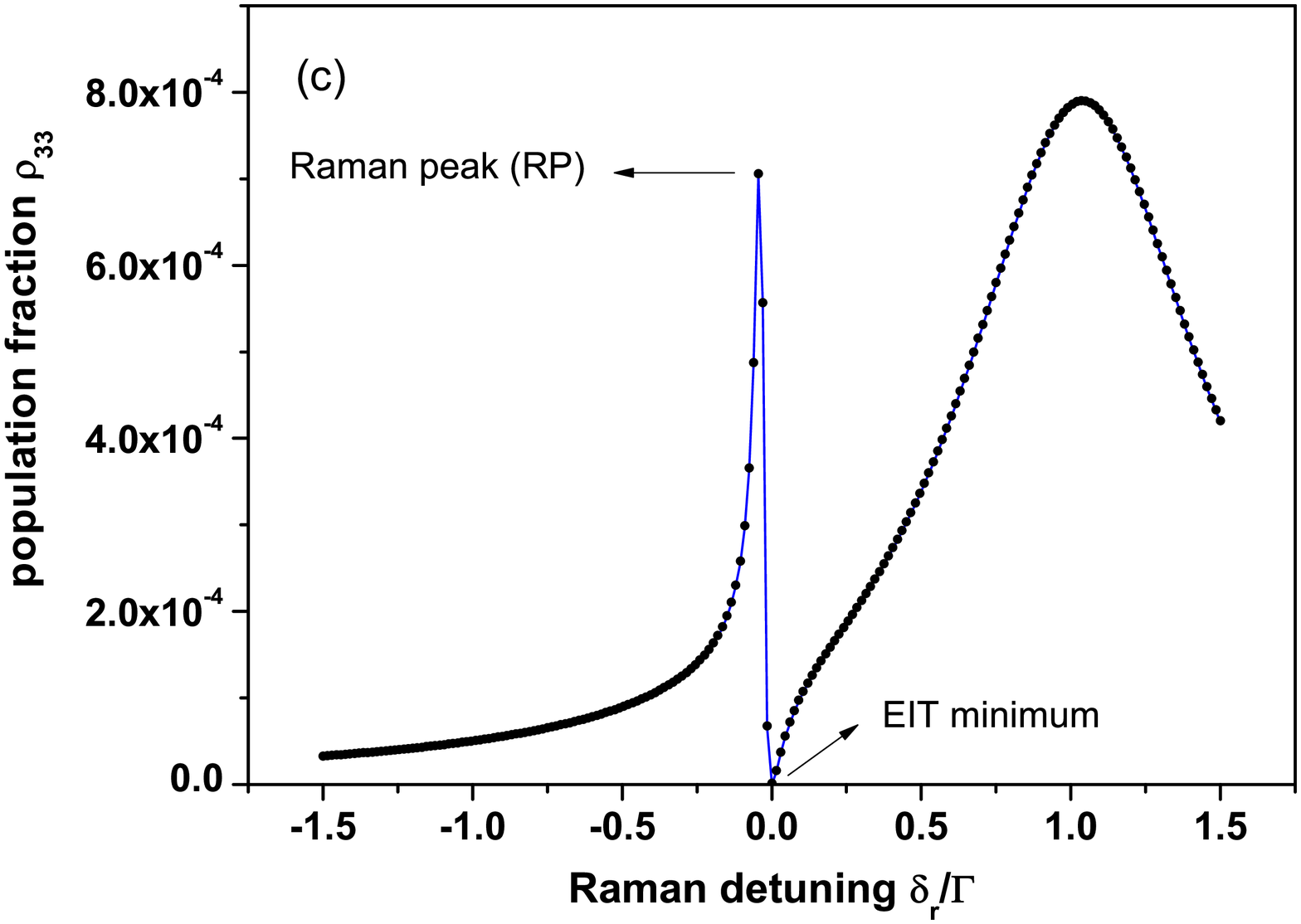}}
\caption{(Color online) Three-level spectra vs $\delta_r$ Raman detuning observed on the $\rho_{33}$ population. For all spectra
$\Gamma_{31}=\Gamma_{32}=\Gamma/2$, $\gamma_c=0$.
 (a) AT spectrum at $\Delta_{0}=0$,  $\Omega_{1}=5\Gamma$,
 and $\Omega_{2}=\Gamma$.  (b) Dark/EIT resonance at $\Delta_{0}=0$,
 $\Omega_{1}=0.2\Gamma$, and  $\Omega_{2}=5\times10^{-2}\Gamma$. (c) FF resonance
 for $\Delta_{0}=\Gamma$, $\Omega_{1}=0.2\Gamma$, $\Omega_{2}=10^{-2}\Gamma$. Solid lines from the analytic solution Eq.~(\ref{fluorescence-solution}) and dots from the numerical integration of Eq.~(\ref{set-Bloch}) at $\tau\gg\tau_{p}(0)$.}
\label{AT-EIT-spectra}
\end{figure}
In examining the steady-state situation with all time derivatives in Eq.~(\ref{set-Bloch})
set to zero, we find the exact expression for the population
of the upper state $\rho_{33}$
\begin{equation}
\rho_{33}=S^{\Lambda}~\frac{(\Delta_{1}-\Delta_{2})^{2}+\gamma_{c}\gamma_{eff}}{(\Delta_{1}-\Delta_{2})[\Delta_{1}-\Delta_{2}-\Delta_{f}]+\Gamma_{eff}^{2}},
\label{fluorescence-solution}
\end{equation}
where $\Delta_{1}-\Delta_{2}=\delta_r$ is the Raman detuning. The expression of
$\Delta_{f}$, the frequency shift affecting the Raman detuning, and
$\Gamma_{eff}$, halfwidth of the two-photon resonance, are reported in the following
subsection. The coherence decay rate is $\gamma_{eff}=\gamma_{c}+\gamma^{*}$ with the
$\gamma^{*}$ saturation rate of the Raman coherence given by
\begin{equation}
\gamma^{*}=\frac{\Omega_{1}^{2}}{\gamma_{2}}+\frac{\Omega_{2}^{2}}{\gamma_{1}},
\label{opticalsaturation}
\end{equation}
Notice that $\gamma_{eff}$ coincides with $\tau_{osc}^{-1}$ for the $\Delta_1=\Delta_2=0$ resonant laser case.\\
\indent $S^{\Lambda}$ of
Eq.~(\ref{fluorescence-solution}) represents the signal amplitude. It contains the broad
features of the $\rho_{33}$ dependence on the optical detunings
$\Delta_{1}, \Delta_{2}$. The fraction on Eq.\eqref{fluorescence-solution},
whose values lie in
the [0,1] interval, determines the $\rho_{33}$ narrow variation
with the Raman detuning $\Delta_{1}-\Delta_{2}$. The three-level signal amplitude
$S^{\Lambda}$ is given by
\begin{equation}
S^{\Lambda}=\frac{S}{1+(3-\frac{\Gamma}{\gamma})S+\frac{\overline{\Omega}_{2}^{2}\Delta_{1}^{2}
+\overline{\Omega}_{1}^{2}\Delta_{2}^{2}}{\gamma_{1}\gamma_{2}}},
\label{saturation-1}
\end{equation}
where we introduced the normalized
dimensionless Rabi frequencies~\cite{note0}
\begin{eqnarray}
\begin{split}
&\overline{\Omega}_{1}^{2}=\Omega_{1}^{2}\frac{\Gamma_{32}\gamma_{1}}{\Omega_{1}^{2}\Gamma_{32}\gamma_{2}+\Omega_{2}^{2}\Gamma_{31}\gamma_{1}},\\
&\overline{\Omega}_{2}^{2}=\Omega_{2}^{2}\frac{\Gamma_{31}\gamma_{2}}{\Omega_{1}^{2}\Gamma_{32}\gamma_{2}+\Omega_{2}^{2}\Gamma_{31}\gamma_{1}}.
\end{split}\label{Rabi-frequencies}
\end{eqnarray}
The saturation parameter $S$ driving
the population exchange between energy levels is determined from
Einstein's rate equations as:
\begin{equation}
S=2\frac{\Omega_{1}^{2}\Omega_{2}^{2}}{(\Gamma_{32}\gamma_{2}\Omega_{1}^{2}+\Gamma_{31}\gamma_{1}\Omega_{2}^{2})}.
\label{saturation-2}
\end{equation}
The imaginary parts
of optical coherences are related to the excited state lineshape
expression by the relation:
\begin{equation}
Im\{\rho_{i3}\}=\frac{\Gamma_{3i}}{2\Omega_{i}}\rho_{33}\hspace{0.5cm}(i=1,2).
\label{opticalcoherences}
\end{equation}
Therefore their lineshape is equivalent to that of $\rho_{33}$.\\
In  Appendix A, we recast all population lineshapes in terms of multi-photon transitions rates, pointing out the light-shift contributions to the optical detuning terms. We verified that the numerical results to be presented in the following can be derived also from that solution.\\
\indent Depending on the  detuning and intensity of the lasers or microwaves driving the
three-level system, the lineshapes associated
to Eqs.~\eqref{fluorescence-solution} and \eqref{opticalcoherences} present very different features, known as
AT spectra, Dark/EIT resonance, and FF profile, associated to different degrees of interference between two-photon transition amplitudes.\\
\indent The AT profile appears
when one Rabi frequency is much larger than the natural linewidth of the excited state
($\Omega_i\gg\Gamma,i=1,2),$ and in addition  $\Omega_1\ll\Omega_2$, or viceversa. Two splitted resonances, a doublet structure, appears in the frequency spectrum, as shown in Fig.~\ref{AT-EIT-spectra}(a).\\
\indent At Rabi frequencies smaller than the excited state width, we reach the DR or EIT configuration where a narrow two-photon resonance is established from the quantum destructive interferences between the transition probability amplitudes \cite{Arimondo:1996,Fleischhauer:2005} as seen in Fig.~\ref{AT-EIT-spectra}(b).
The system is placed in the dark state uncoupled from the driving fields. Note that at exact resonance, $\Delta_1=\Delta_2=0$,
$\rho_{33}=0$ when $\gamma_c=0$. These regimes are  characterized $(\Delta_1\approx\Delta_2\lesssim\Gamma)$, with $\Omega_1\approx\Omega_2$ for the Dark Resonance, and $\Omega_i\gg\Omega_j; \Omega_i, \Omega_j<\Gamma$ for the EIT resonance.\\
\indent The FF lineshape plotted in Fig.~\ref{AT-EIT-spectra}(c) is originated when one Rabi frequency is much larger than the second one, and in presence of an optical detuning from the excited state($\Omega_i\gg\Omega_j$, $\Delta_1,\Delta_2\geq\Gamma$). Two resonances appear in the FF spectrum, one broad corresponding to the saturated one-photon resonance. The second sharp feature exhibits a characteristics asymmetric response, highly sensitive to changes in the
system parameters and centered around the $\delta_r \approx 0$ Raman detuning. Its minimum
is associated to the DR, or EIT dip, while the maximum is the Raman peak, or bright resonance, associated to the preparation of the coherent superposition of $|1>$ and $|2>$ states coupled to the driving electromagnetic fields. Both the EIT dip and the Raman peak are manifestations of the interference between the one-photon and two-photon amplitudes~\cite{Lounis:1992}. The asymmetry of the FF profile is reversed by changing the relative ratio between the Rabi frequencies.
\begin{figure*}[!!t]
\resizebox{8.5cm}{!}{\includegraphics{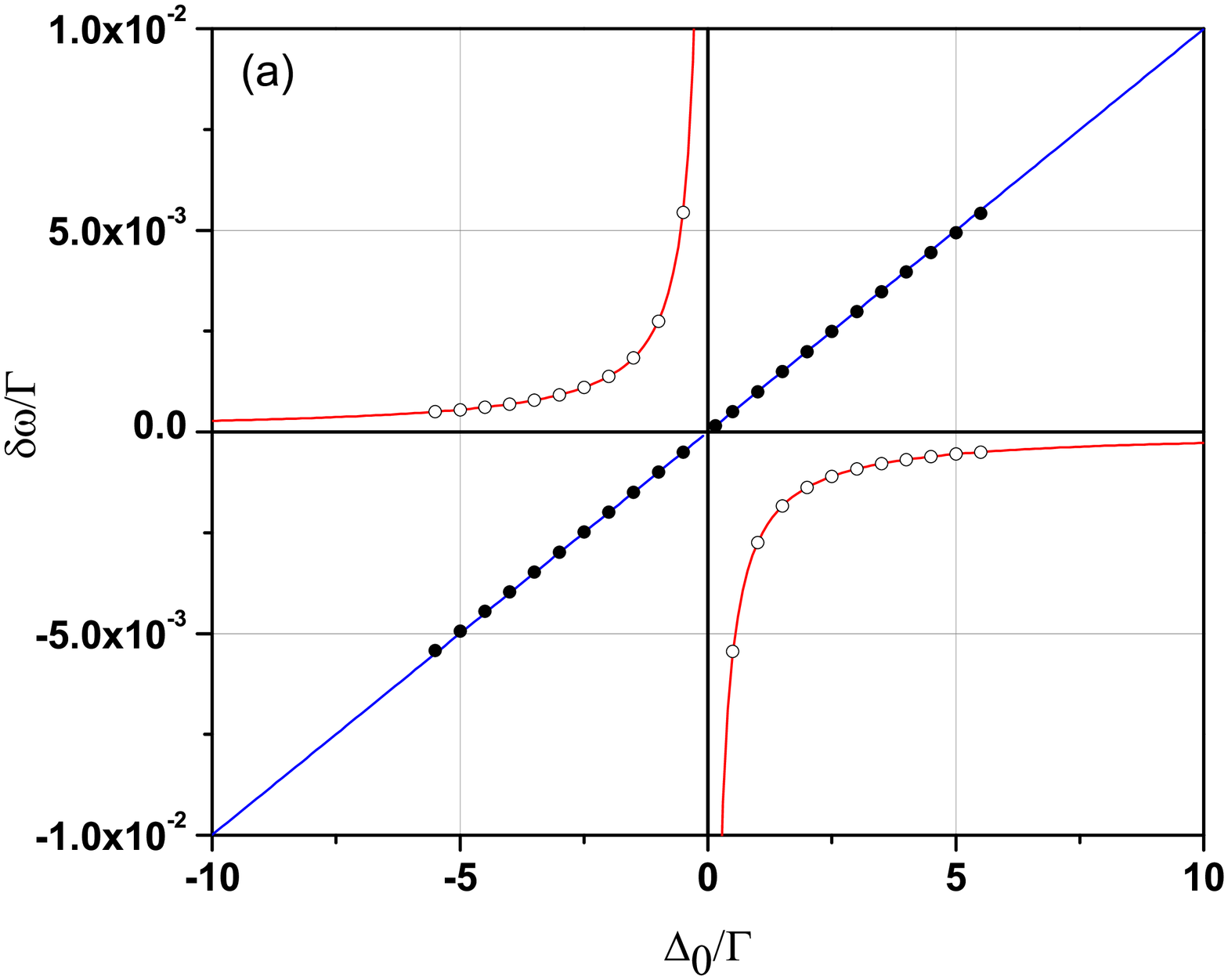}}
\resizebox{8.5cm}{!}{\includegraphics{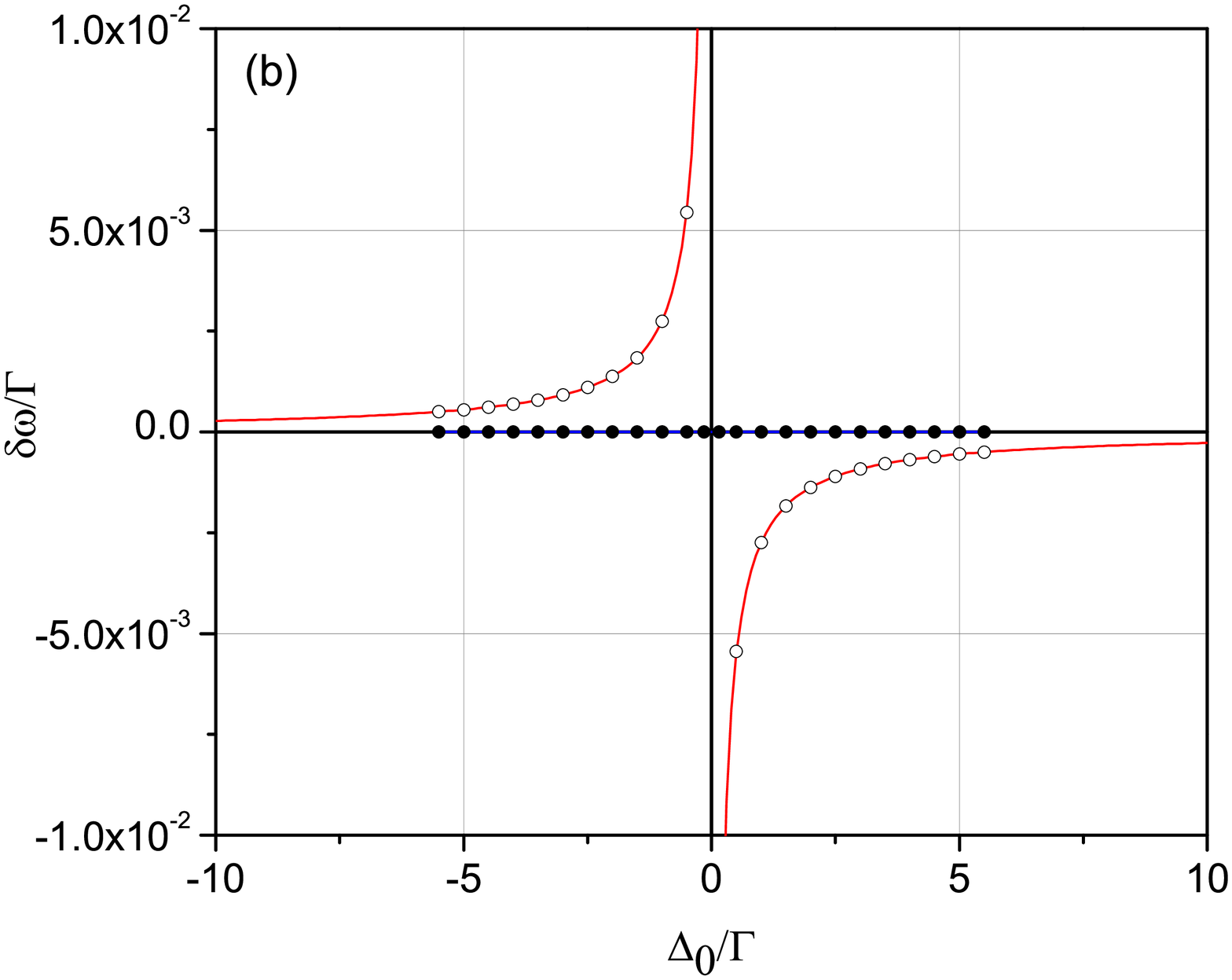}}
\caption{(Color online) Frequency shifts of the Dark/EIT dip and Raman peak observed on the $\rho_{33}$ excited state population  versus the optical detuning $\Delta_{0}$  from Eq.~\eqref{shift}(solid blue and red lines, respectively), and from the numerical integration of Eq.~(\ref{set-Bloch}) (solid dots $\bullet$ and open dots $\circ$, respectively).
Parameters are $\Gamma_{31}=\Gamma_{32}=\Gamma/2$, $\Omega_{1}=5\times10^{-2}\Gamma$, and $\Omega_{2}=10^{-3}\Gamma$. In (a)  $\gamma_{c}=5\times10^{-4}\Gamma$; in (b) $\gamma_{c}=0$ leads to a vanishing shift for the EIT resonance. }
\label{cw-shift}
\end{figure*}

\subsection{$\Gamma_{eff}$ Raman linewidth}

The sub-natural EIT resonance of Fig.~\ref{AT-EIT-spectra}(b) experiences a
linewidth which is power broadened by
the optical saturation rate $\gamma^{*}$ of Eq.~\eqref{opticalsaturation}. Let's note that such
power broadening is important even for a laser intensity where the saturation is negligible on the optical
transitions. In fact ref.~\cite{Kocharovskaya:1990} introduced a coherence saturation intensity,
defined by $\gamma^{*}=\gamma_c$, smaller than the optical saturation intensity. The exact expression of the $\Gamma_{eff}$
Raman halfwidth is
\begin{equation}
\Gamma_{eff}=\sqrt{\frac{\left(\gamma_{eff}+3\gamma_{c}S\right)\gamma_{eff}
}{1+(3-\frac{\Gamma}{\gamma})S+\frac{\overline{\Omega}_{2}^{2}\Delta_{1}^{2}
+\overline{\Omega}_{1}^{2}\Delta_{2}^{2}}{\gamma_{1}\gamma_{2}}}}\times{\sqrt{1+\zeta}},
\label{Gammaeff}
\end{equation}
where the factor $\zeta$ is
\begin{equation}
\begin{split}
\zeta=&\frac{\Gamma\frac{\gamma_{c}S}{2\gamma_{1}\gamma_{2}}\left[2\Delta_{1}\Delta_{2}+\Delta_{1}^{2}-\Delta_{2}^{2}\right]
+\gamma_{c}^{2}\frac{\overline{\Omega}_{2}^{2}\Delta_{1}^{2}
+\overline{\Omega}_{1}^{2}\Delta_{2}^{2}}{\gamma_{1}\gamma_{2}}}{(\gamma_{eff}+3\gamma_{c}S)\gamma_{eff}}\\
&+\frac{\frac{(\gamma_{1}-\gamma_{2})S}{2\gamma_{1}\gamma_{2}}\left(\Gamma_{32}\gamma_{2}-\Gamma_{31}\gamma_{1}\right)}
{\gamma_{eff}+3\gamma_{c}S}.
\end{split}
\end{equation}
$\zeta$ is very small ($\zeta\ll1$) in a quasi-resonant laser interaction (according to the condition $\dfrac{\gamma_{c}}{\Gamma}\Delta_{1}^2,\dfrac{\gamma_{c}}{\Gamma}\Delta_{2}^2\ll\Omega^2$) and a pure radiative process. As long as $\gamma_c, \Omega_1,\Omega_2\ll\Gamma$, $\Gamma_{eff}$ determines the half-linewidth of the sub-natural resonance. In that regime and for the pure radiative case, $\Gamma_{eff}$ is well approximated by $\tau_{osc}^{-1}$.

\subsection{$\Delta_{f}$ Raman frequency-shift }
The frequency shift
$\Delta_{f}$ correcting the $\delta_r=0$ Raman detuning condition is given by
\begin{equation}
\Delta_{f}=\Delta_{LS}+\Delta_{DS},
\end{equation}
with the $\Delta_{LS}$ light-shift (LS) expression including the saturation effect given by
\begin{eqnarray}
\begin{split}
\Delta_{LS}&=\frac{\frac{2}{\gamma_{1}\gamma_{2}}\left[\Omega_{2}^{2}\overline{\Omega}_{2}^{2}\Delta_{1}
-\Omega_{1}^{2}\overline{\Omega}_{1}^{2}\Delta_{2}\right]}
{1+(3-\frac{\Gamma}{\gamma})S+\frac{\overline{\Omega}_{2}^{2}\Delta_{1}^{2}
+\overline{\Omega}_{1}^{2}\Delta_{2}^{2}}{\gamma_{1}\gamma_{2}}}\\
&-\frac{(\gamma_{1}-\gamma_{2})}{2\gamma_{1}\gamma_{2}\gamma}\frac{
\Gamma S \left[\gamma_{2}\Delta_{1}+\gamma_{1}\Delta_{2}\right]}{{1+(3-\frac{\Gamma}{\gamma})S+\frac{\overline{\Omega}_{2}^{2}\Delta_{1}^{2}
+\overline{\Omega}_{1}^{2}\Delta_{2}^{2}}{\gamma_{1}\gamma_{2}}}}.
\end{split}\label{LS}
\end{eqnarray}
Notice that the second term of the above expression vanishes for the symmetric $\Lambda$ scheme, i.e.,
with $\gamma_1=\gamma_2$ (pure radiative process). In that case, the first term of $\Delta_{LS}$ could be associated to the light-shift expression, as pointed out in~\cite{Brewer:1975}. The $\Delta_{DS}$ decoherence shift (DS) depends on the $\gamma_c$ rate as
\begin{equation}
\Delta_{DS}=\gamma_c\frac{S}{2\gamma_{1}\gamma_{2}}\frac{\left[\Gamma(\Delta_{1}+\Delta_{2})-(\Gamma_{31}\Delta_{1}-\Gamma_{32}\Delta_{2})\right]}
{{1+(3-\frac{\Gamma}{\gamma})S+\frac{\overline{\Omega}_{2}^{2}\Delta_{1}^{2}
+\overline{\Omega}_{1}^{2}\Delta_{2}^{2}}{\gamma_{1}\gamma_{2}}}}.
\label{deplacement-collisionnel}
\end{equation}
Let us emphasize that $\Delta_f$ is always null at $\Delta_1=\Delta_2=0$ resonance~\cite{note}.

\subsection{Approximated frequency-shifts of EIT and FF resonances}

Instead of the previous Section exact expression correcting the Raman detuning condition in the denominator, it is useful to derive the effective shift of the Dark/EIT resonance minimum, which is very relevant for precision spectroscopy or clock resonance. The calculation of the DR/EIT and FF frequency shifts requires to examine
Eq.~(\ref{fluorescence-solution}) with the Raman detuning $\delta_{r}$ as a free parameter.
A valid approximation for the EIT and Raman-peak shifts in various excitation configurations can be found when optical detuning
$\Delta_{1}\sim\Delta_{2}\sim\Delta_{0}$ are tuned around the Raman condition
$\Delta_{1}-\Delta_{2}=\delta_{r}$.
A differentiation of Eq.~(\ref{fluorescence-solution})
versus the $\delta_{r}$ parameter leads to roots of a quadratic equation defining the following extrema $\delta\omega_{33}(\Delta_{0})$ of the EIT/FF lineshapes:
\begin{eqnarray}
\begin{split}
&\delta\omega_{33}(\Delta_{0})\approx\\
&\frac{\Gamma_{eff}^{2}-\gamma_{c}\gamma_{eff}}{\Delta_{f}}\left(1\mp\sqrt{1+\frac{\gamma_{c}\gamma_{eff}\Delta_{f}^{2}}
{(\Gamma_{eff}^{2}-\gamma_{c}\gamma_{eff})^{2}}}\right).
\end{split}\label{shift}
\end{eqnarray}
The $\mp$ solutions refer to the extrema of the FF
lineshape. The minus (plus) sign holds for EIT dip (Raman Peak) when
$\gamma_{c}\gamma_{eff}/\Gamma_{eff}^{2}<1$, and the opposite when $\gamma_{c}\gamma_{eff}/\Gamma_{eff}^{2}>1$.
\begin{figure}[!t]
\centering
\hfill\resizebox{8.5cm}{!}{\includegraphics{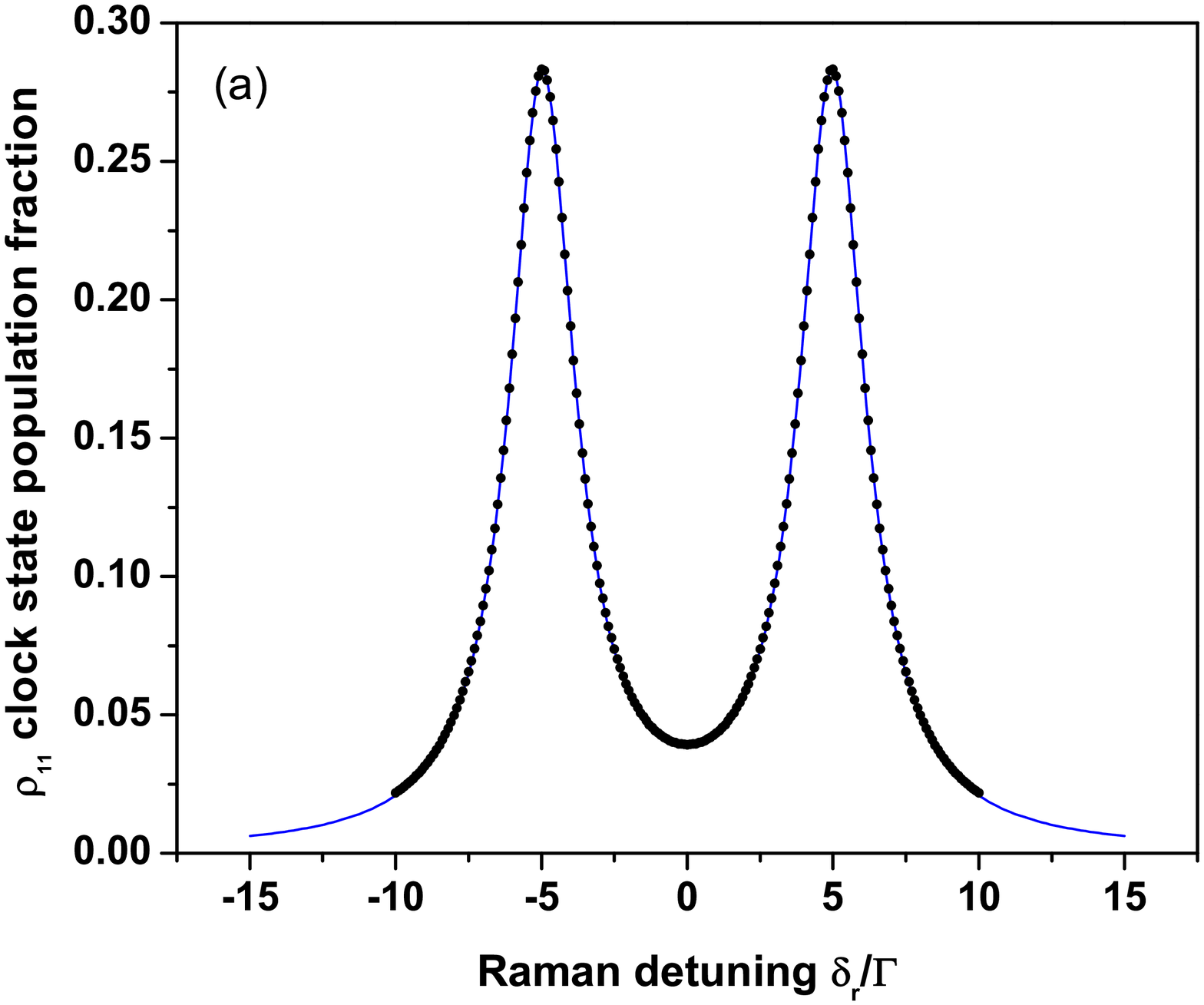}}
\hfill\resizebox{8.5cm}{!}{\includegraphics{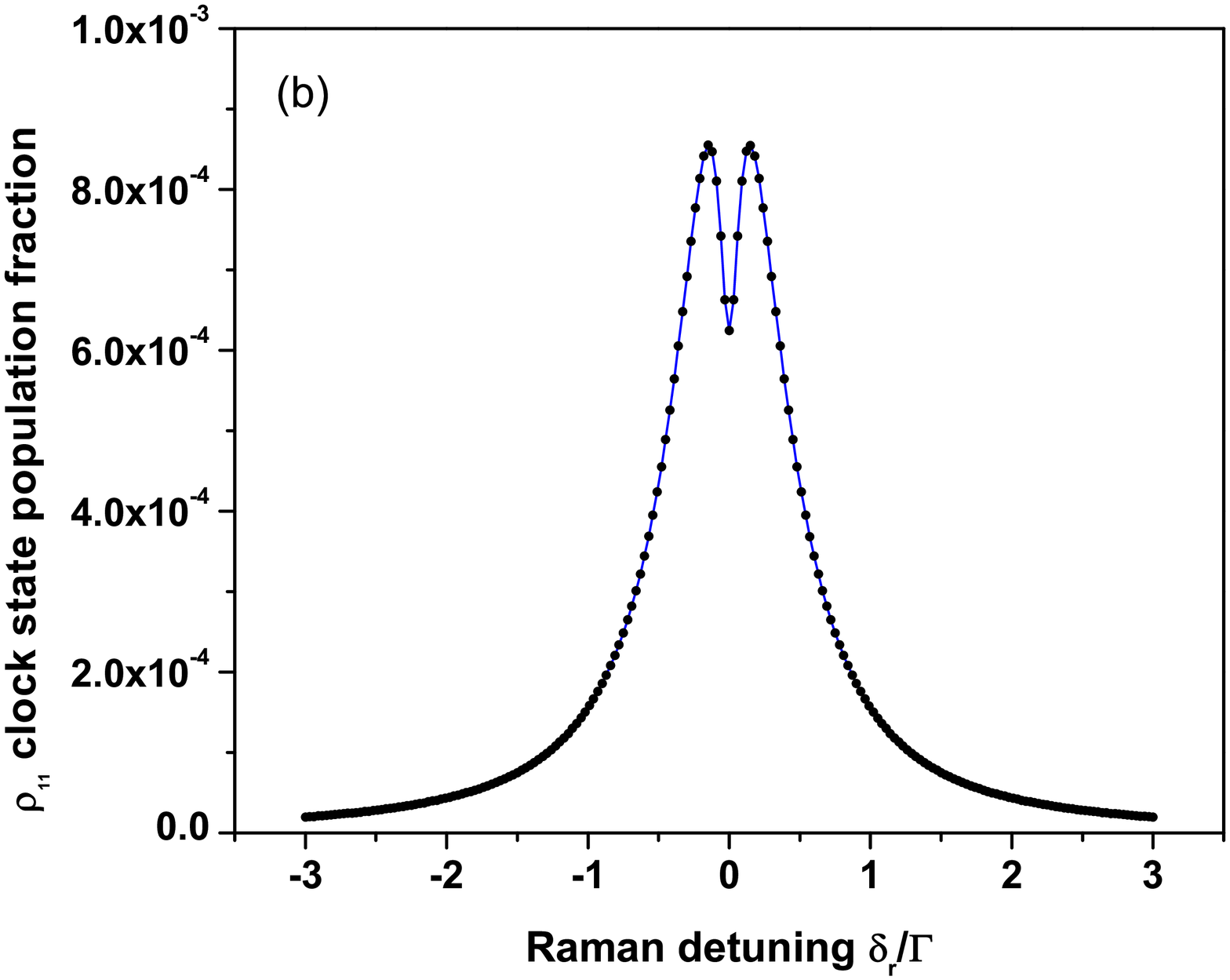}}
\hfill\resizebox{8.5cm}{!}{\includegraphics{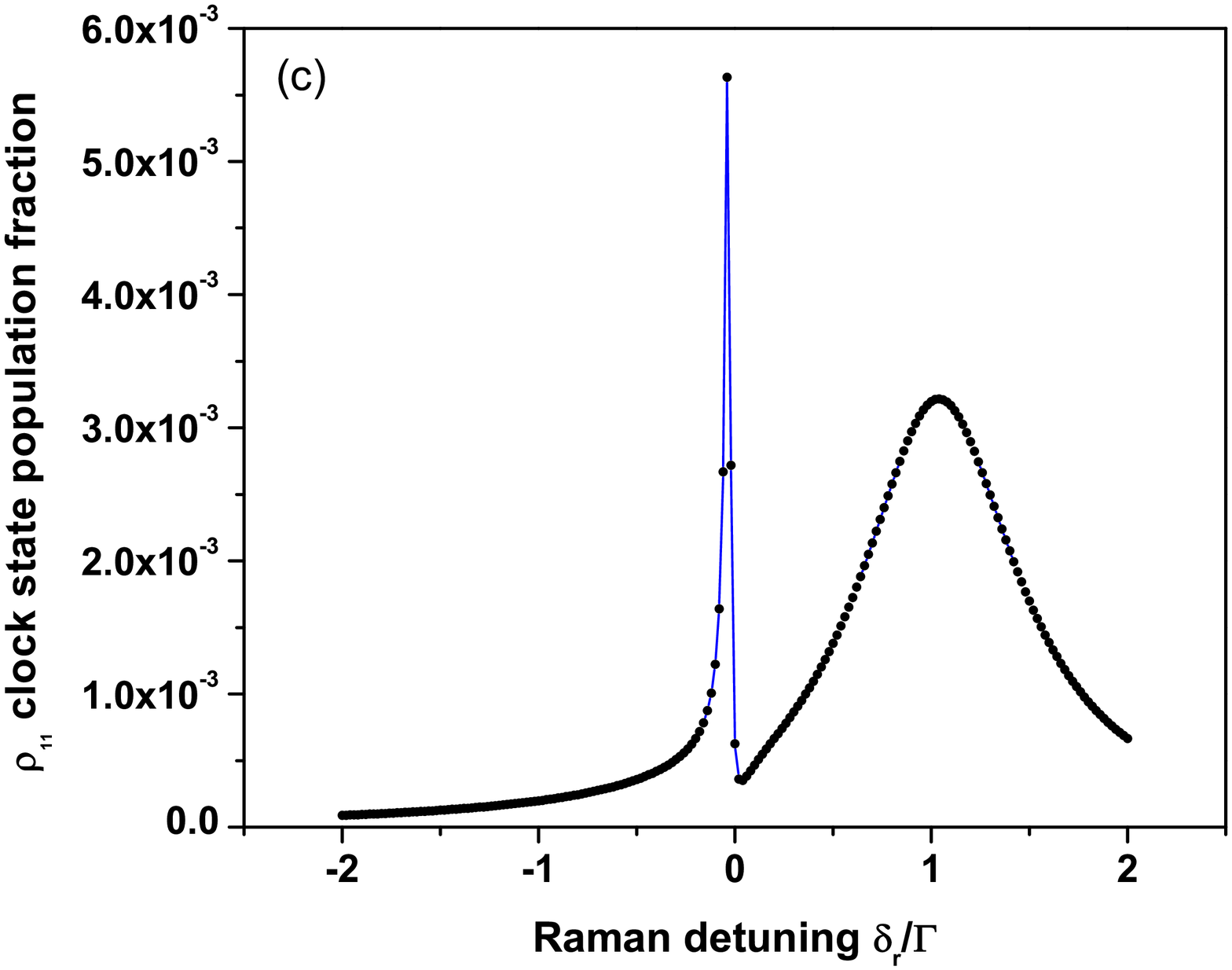}}
\caption{(Color online) Three-level spectra versus the $\delta_r$ Raman detuning observed on the $\rho_{11}$ population using Eq.~(\ref{populations}), solid blue line, and the numerical
integration of Eq.~(\ref{set-Bloch}), solid dots ($\bullet$).  In (a) AT resonance, in (b) Lamb dip lineshape, and in  (c) FF resonance. System parameters as in Fig. 2. In (a) and (b) Rabi frequencies as in Fig .2. In (c)
$\Omega_{1}=0.2\Gamma$, $\Omega_{2}=5\times10^{-3}\Gamma$, and $\Delta_{0}=\Gamma$. }\label{population}
\end{figure}
\begin{figure}[!b]
\centering
\hfill\resizebox{8.5cm}{!}{\includegraphics{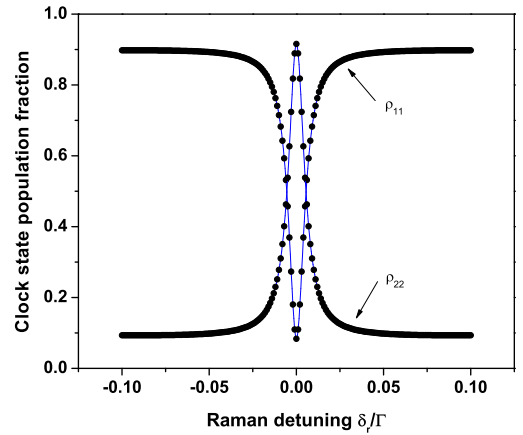}}
\caption{(Color online) Population inversion resonance between the $|1>$ and $|2>$ states, monitored on  $\rho_{11}$ and $\rho_{22}$,
for unbalanced decay rates, $\Gamma_{32}=10^{-5}\Gamma$, $\Gamma_{31}=\Gamma-\Gamma_{32}$.
 Solid dots ($\bullet$) are the result of the numerical
integration of Eq.~(\ref{set-Bloch}). Other parameters are $\Delta_0=0$, $\gamma_{c}=0$, $\Omega_{1}=5\times10^{-2}\Gamma$ and $\Omega_{2}=5\times 10^{-4}\Gamma$.}
\label{population-inversion}
\end{figure}
\begin{figure*}[!!t]
\resizebox{8.5cm}{!}{\includegraphics[angle=0]{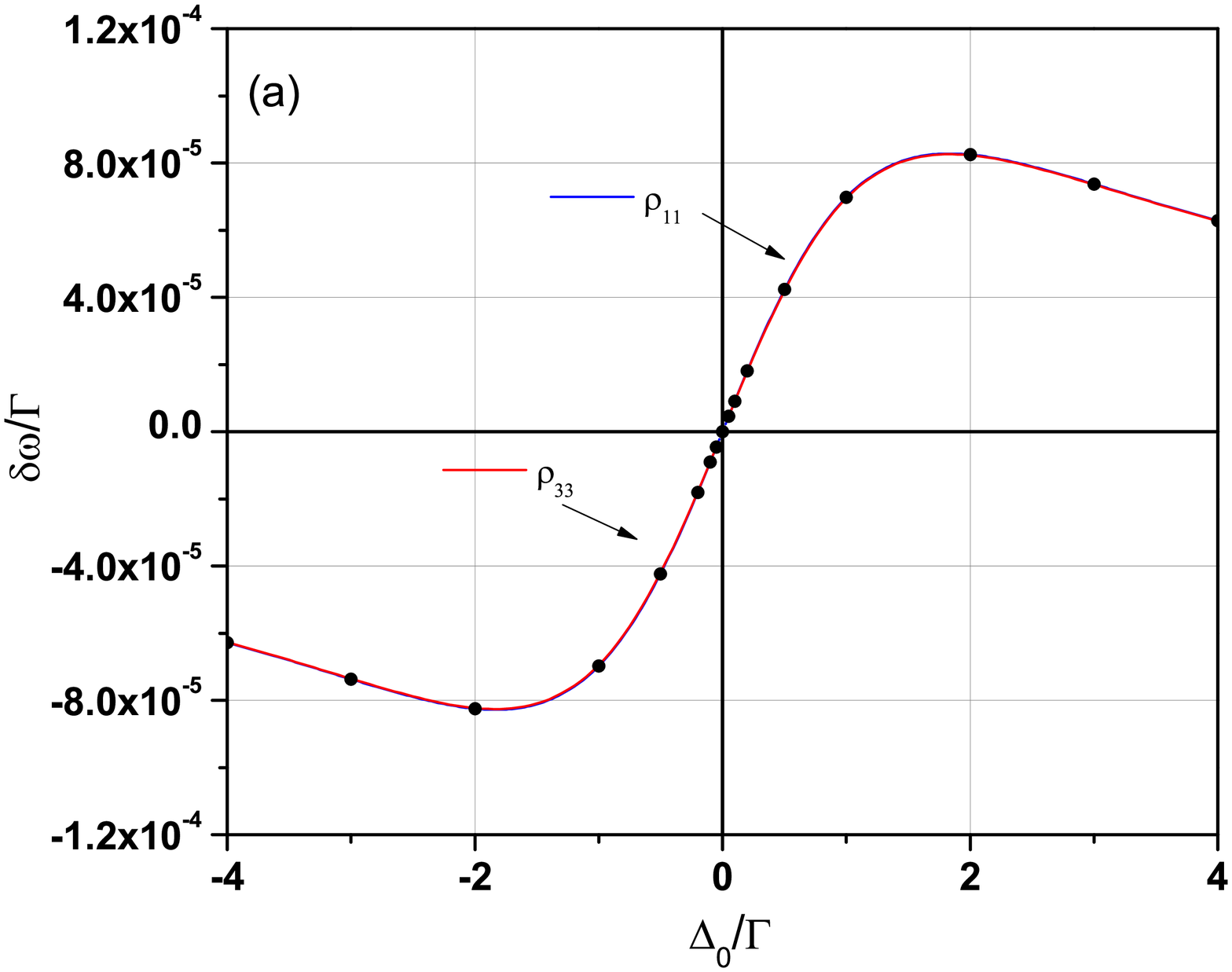}}
\resizebox{8.5cm}{!}{\includegraphics[angle=0]{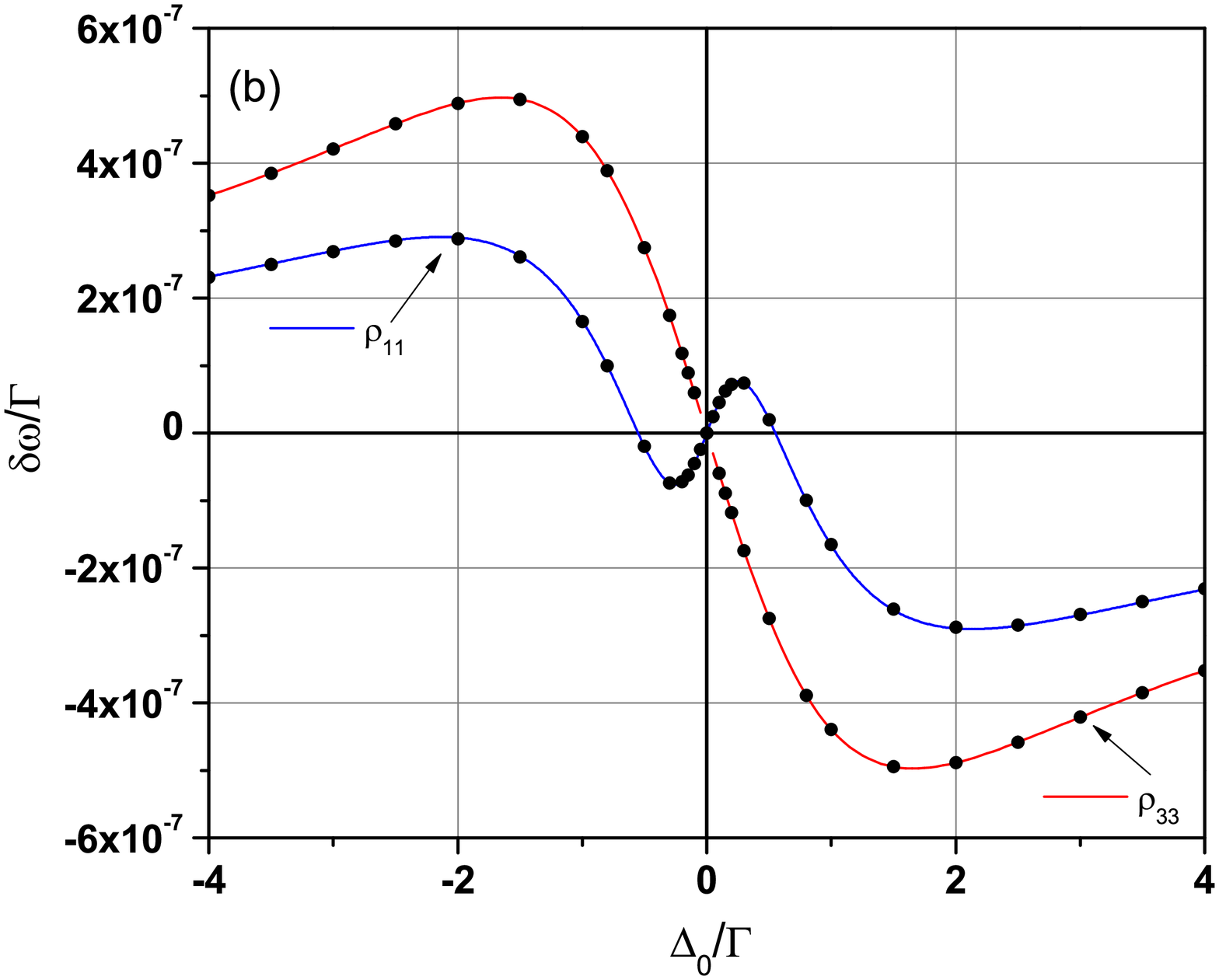}}
\caption{(Color online) Frequency shift of the population inversion resonance observed on $\rho_{11}$(or $\rho_{22}$) as derived from
Eq.~(\ref{population-shift11}) (solid blue line) and frequency-shift of the $\rho_{33}$ EIT resonance  from Eq.~(\ref{shift}) (solid red line) versus the optical detuning $\Delta_{0}$, for unbalanced decay rates. Solid dots ($\bullet$) are from the numerical integration of Eq.~(\ref{set-Bloch}) with parameters  $\Omega_{1}=5\times10^{-2}\Gamma$, $\gamma_{c}=5~10^{-4}\Gamma$ and $\Omega_{2}=5\times 10^{-4}\Gamma$. In (a)  $\Gamma_{32}=10^{-5}\Gamma$, $\Gamma_{31}=\Gamma-\Gamma_{32}$. Note that the  $\delta\omega_{33}$ shift is indistinguishable from the  $\delta\omega_{11}$ shift. In (b) $\Gamma_{32}=10^{-10}\Gamma$, $\Gamma_{31}=\Gamma-\Gamma_{32}$.}
\label{shift-population}
\end{figure*}
Fig.~\ref{cw-shift} shows the continuous-wave frequency shift versus the common optical detuning $\Delta_0$ using Eq.~(\ref{shift}).
Results for both $\gamma_c=5.10^{-4}\Gamma$ and $\gamma_c=0$  are presented.
The EIT decoherence shift, proportional to $\gamma_c$ exhibits a linear dependence on the optical detuning when radiative decay terms are symmetrical. As pointed in~\cite{Orriols:1979}, when the decoherence term vanishes ($\gamma_{c}=0$),
there is no shift of the EIT minimum~\cite{note2} except if we take into account external small off-resonant level contributions \cite{Declercq:1983}.
The Raman peak shift produced by light and decoherence varies with the inverse of the optical detuning as discussed by \cite{Stalgies:1998} and observed experimentally in \cite{Siemers:1992}.
For a quasi-resonant interaction we can further simplify the above expression for the Raman Peak and the EIT dip when $\gamma_{c}\gamma_{eff}/\Gamma_{eff}^{2}\ll1$ and $\Delta_{f}^{2}/\Gamma_{eff}^{2}\ll1$. Near the two-photon resonance, the shifts of Eq.~(\ref{shift}) can be accurately approximated as
\begin{eqnarray}
\begin{split}
\delta\omega_{33}^{EIT}(\Delta_{0})\approx&-\gamma_{c}\Delta_{f}\frac{\gamma_{eff}}{2\Gamma_{eff}^{2}},\\
\delta\omega_{33}^{RP}(\Delta_{0})\approx&2\frac{\Gamma_{eff}^{2}}{\Delta_{f}}.
\end{split}\label{Raman-shift}
\end{eqnarray}
The EIT dip frequency-shift is thus roughly given by the product of the $\gamma_{c}$ decoherence rate and the $\Delta_{f}$ Raman frequency shift divided by the $\Gamma_{eff}$ linewidth of the sub-natural resonance. Such a dependence was pointed out by~\cite{Santra:2005} and by \cite{Wynands:1999} based on a theoretical analysis of~\cite{Fleischhauer:1994}, and as mentioned in~\cite{MacDonnell:2004}, was earlier derived in~\cite{Kofman:1997}.

\section{Steady-state lineshapes of clock state populations $\rho_{11},\rho_{22}$}

\subsection{The two-photon resonance}

We focus now on clock-state resonances observed on the $\rho_{11}$, $\rho_{22}$ populations and linked to the Raman coherence between those states. The clock state populations may be expressed in an exact form similar to that of Eq.~\eqref{fluorescence-solution} as
\begin{eqnarray}
\begin{split}
\rho_{11}&=S^{11}\frac{(\Delta_{1}-\Delta_{2})[\Delta_{1}-\Delta_{2}-\Delta_{11}]+\gamma_{11}^{2}}{(\Delta_{1}-\Delta_{2})[\Delta_{1}-\Delta_{2}-\Delta_{f}]+\Gamma_{eff}^{2}},\\
\rho_{22}&=S^{22}\frac{(\Delta_{1}-\Delta_{2})[\Delta_{1}-\Delta_{2}+\Delta_{22}]+\gamma_{22}^{2}}{(\Delta_{1}-\Delta_{2})[\Delta_{1}-\Delta_{2}-\Delta_{f}]+\Gamma_{eff}^{2}}. \label{populations}
\end{split}
\end{eqnarray}
where
\begin{eqnarray}
\begin{split}
S^{ii}&=\left(1+\frac{\widetilde{\gamma}_{i}\Gamma_{3i}}{2\Omega_{i}^{2}}\right) S^{\Lambda}, (i=1,2),\\
\Delta_{11}&=\overline{\gamma}_{11}\left(\Delta_{1}+\gamma_{1}\overline{\Delta}\right),\\
\Delta_{22}&=\overline{\gamma}_{22}\left(\Delta_{2}-\gamma_{2}\overline{\Delta}\right),\\
\overline{\Delta}&=\frac{\Omega_{2}^{2}\Gamma_{31}\gamma_{2}\Delta_{1}-\Omega_{1}^{2}\Gamma_{32}\gamma_{1}\Delta_{2}}
{\gamma_{1}\gamma_{2}(\Omega_{2}^{2}\Gamma_{31}+\Omega_{1}^{2}\Gamma_{32})},
\end{split}\label{}
\end{eqnarray}
and
\begin{eqnarray}
\begin{split}
\gamma_{11}^{2}&=\gamma_{c}\gamma_{eff}+\overline{\gamma}_{11}\left(\gamma_{1}\gamma_{eff}-\gamma_{c}\Delta_{1}\overline{\Delta}\right),\\
\gamma_{22}^{2}&=\gamma_{c}\gamma_{eff}+\overline{\gamma}_{22}\left(\gamma_{2}\gamma_{eff}+\gamma_{c}\Delta_{2}\overline{\Delta}\right),
\end{split}\label{}
\end{eqnarray}
with
\begin{eqnarray}
\begin{split}
\overline{\gamma}_{11}&=\frac{\Gamma_{31}\Omega_{2}^{2}+\Gamma_{32}\Omega_{1}^{2}}{2\Omega_{1}^{2}\gamma_{1}+(\Delta_{1}^{2}+\gamma_{1}^{2})\Gamma_{31}},\\
\overline{\gamma}_{22}&=\frac{\Gamma_{31}\Omega_{2}^{2}+\Gamma_{32}\Omega_{1}^{2}}{2\Omega_{2}^{2}\gamma_{2}+(\Delta_{2}^{2}+\gamma_{2}^{2})\Gamma_{32}}.
\end{split}\label{}
\end{eqnarray}
Fig.~\ref{population} reports the population resonance under various saturation conditions, matching the AT, EIT and FF lineshapes of Fig.~\ref{AT-EIT-spectra}. Notice that the EIT regime of Fig.~\ref{population}(b) corresponds to the case of Rabi frequencies smaller than the natural decay rate of the excited state. The "Lamb-dip" like lineshape for the resulting quasi saturated transition can conveniently be observed in the three-level configuration. This method of spectroscopy without Doppler broadening was proposed and experimentally accomplished by Javan and Schlossberg in \cite{Schlossberg:1966,Schlossberg-bis:1966}. In such situation, the dip can be narrower than the homogeneous linewidth of the population resonance as in Fig.~\ref{population}(b).\\
\indent The clock state populations depend strongly on the normalized branching ratio difference $\Upsilon$ and on the Rabi frequencies driving atomic or molecular transitions. A numerical analysis of~\cite{Jyotsna:1995} demonstrated the occurrence of a strong population transfer for unequal $\Gamma_{31},\Gamma_{32}$ decay rates.
Fig.~\ref{population-inversion} shows the steady-state complete population transfer for $\Omega_{1}\gg\Omega_{2}$ and $\Upsilon\approx 1$ using the asymmetric decay rates associated to an alkaline-earth three-levels system as strontium atoms~\cite{Santra:2005,Zanon-Willette:2006}.
A large coherent population transfer $\rho_{22}-\rho_{11}=\pm1$ is achieved when $\Omega_{1}\gg\Omega_{2}$ for $\Upsilon\sim+1$ or $\Omega_{2}\gg\Omega_{1}$
for $\Upsilon\sim-1$.

\subsection{Approximated frequency-shift of the two-photon resonance}

In an alkaline-earth frequency clock probing scheme, the large population transfer regime of Fig.~\ref{population-inversion} may be used to detect one lower state population, or the population difference between clock states. Thus, it is important to derive the two-photon shift also in this scheme. An analysis equivalent to the derivation of Eq.~(\ref{shift}) when $\Delta_{1}\sim\Delta_{2}\sim\Delta_{0}$, we obtain, the frequency shift of $\rho_{11}$. For $\rho_{22}$, we make a similar derivation also using the population conservation condition. We obtain the following expression for $\delta\omega_{11}(\Delta_{0})$ and similarly for $\delta\omega_{22}(\Delta_{0})$
\begin{equation}
\begin{split}
&\delta\omega_{11}(\Delta_{0})\sim\delta\omega_{22}(\Delta_{0})\approx\\
&\frac{\gamma_{11}^{2}-\Gamma_{eff}^{2}}{\Delta_{11}-\Delta_{f}}\left(1\mp\sqrt{1+\frac{(\Delta_{f}-\Delta_{11})(\gamma_{11}^{2}\Delta_{f}-\Delta_{11}\Gamma_{eff}^{2})}{(\Gamma_{eff}^{2}-\gamma_{11}^{2})^{2}}}\right).
\end{split}\label{population-shift11}
\end{equation}
For the specific radiative configuration of alkaline-earth species shown in Fig.~\ref{population-inversion}, only the solution with the minus sign is needed but $\mp$ solutions generally refer to the extrema of a dispersive lineshape.
Our standard choice of the laser detunings $\Delta_{1}=\Delta_0$ and $\Delta_{2}=\Delta_0-\delta_r$ introduces a very small difference in expressions of the frequency-shifts affecting each clock state population.
For a quasi-resonant interaction when $\gamma_{c}\gamma_{eff}/\Gamma_{eff}^{2}\ll1$ with $\Delta_{f}/\Gamma_{eff}\ll1$ and $\Delta_{11}/\Gamma_{eff}\ll1$, the expression may be simplified to yield:
\begin{equation}
\delta\omega_{11}(\Delta_{0})\sim\delta\omega_{22}(\Delta_{0})\approx\frac{\Delta_{11}-\frac{\gamma_{11}^{2}}{\Gamma_{eff}^{2}}\Delta_{f}}{2(1-\frac{\gamma_{11}^{2}}{\Gamma_{eff}^{2}})}.
\label{eq26}
\end{equation}
The frequency shift versus the common mode optical detuning $\Delta_{0}$ affecting the $\rho_{11}$(equivalently $\rho_{22}$) resonance is plotted in Fig.~\ref{shift-population}(a) and (b) for two particular ratios of the relaxation rates by spontaneous emission. In both cases, the shift of the two-photon resonance measured on the $\rho_{11}$ or $\rho_{22}$ observables has a dispersive lineshape versus the optical detuning $\Delta_{0}$. The slope is completely reversed owing to a nonlinear behavior when the ratio $\Upsilon\mapsto1$ as in Fig.~\ref{shift-population}(b). A comparison with the frequency-shift of the excited state $\rho_{33}$ is also included in the figure. Notice the difference in the $\rho_{11}$/$\rho_{33}$ shifts for the case of a large asymmetry in the spontaneous decay rates.

\section{Steady-state lineshape of Raman coherence $\rho_{12}$}

\begin{figure}[!!t]
\centering
\hfill\resizebox{8.5cm}{!}{\includegraphics{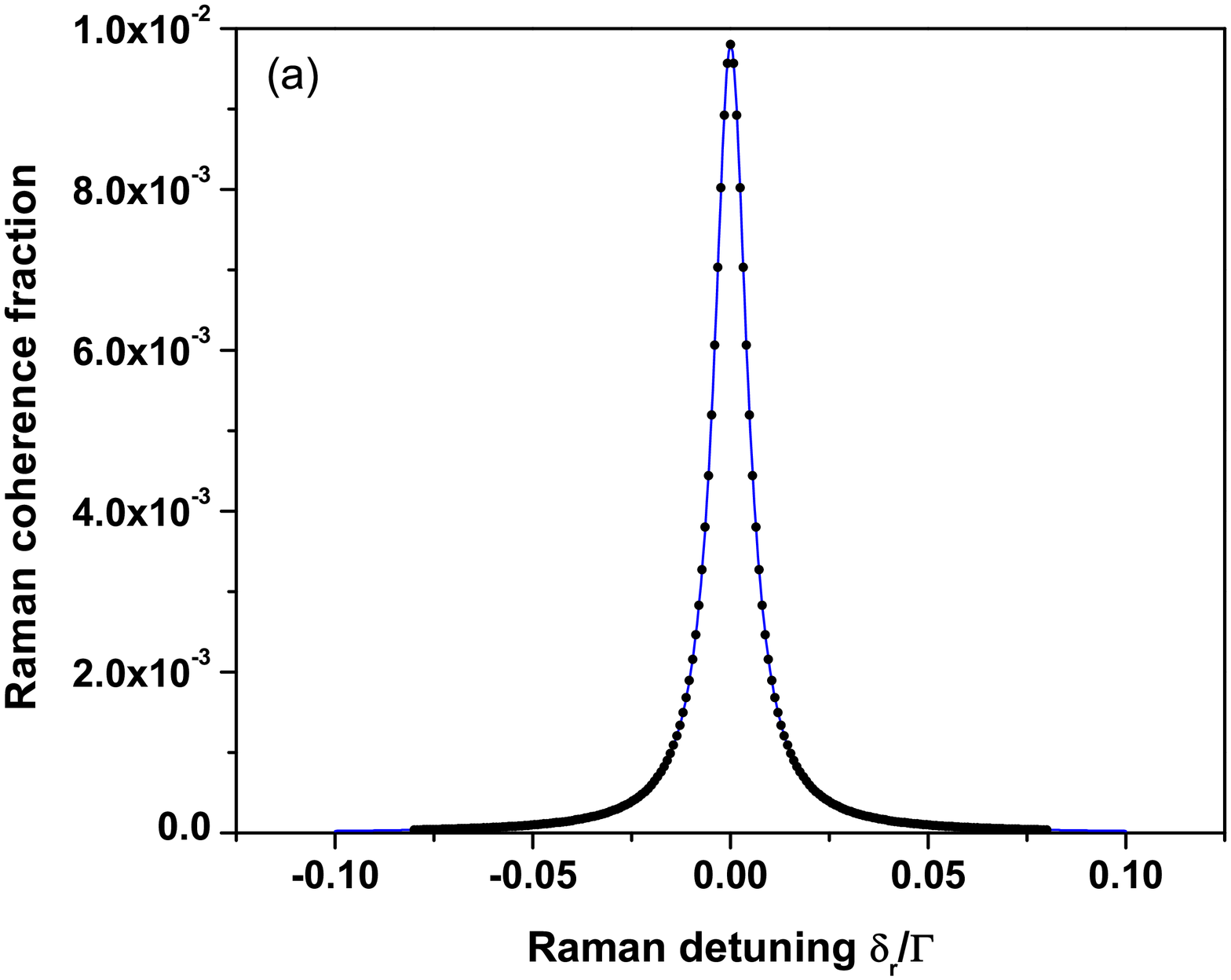}}
\hfill\resizebox{8.5cm}{!}{\includegraphics{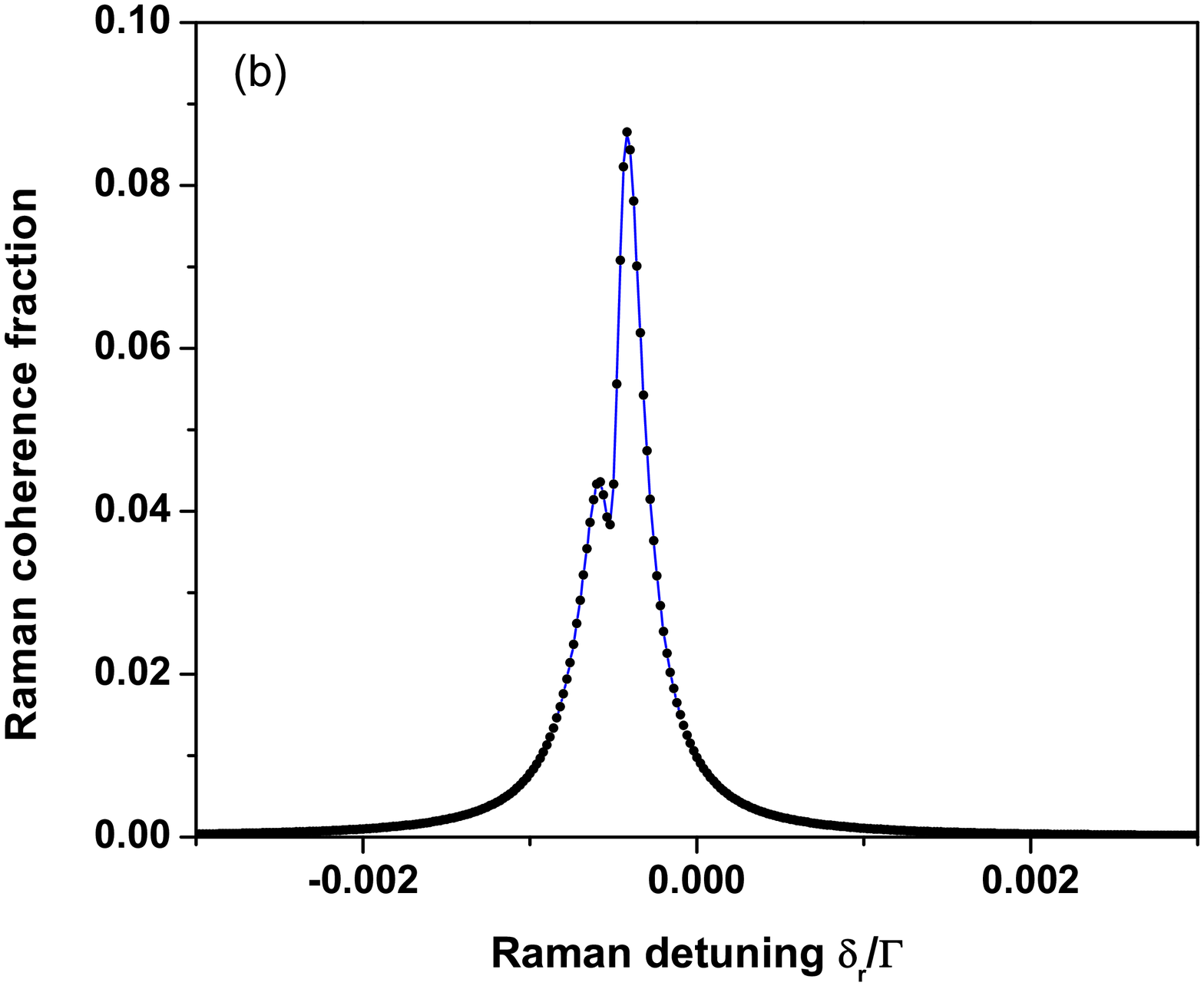}}
\hfill\resizebox{8.5cm}{!}{\includegraphics{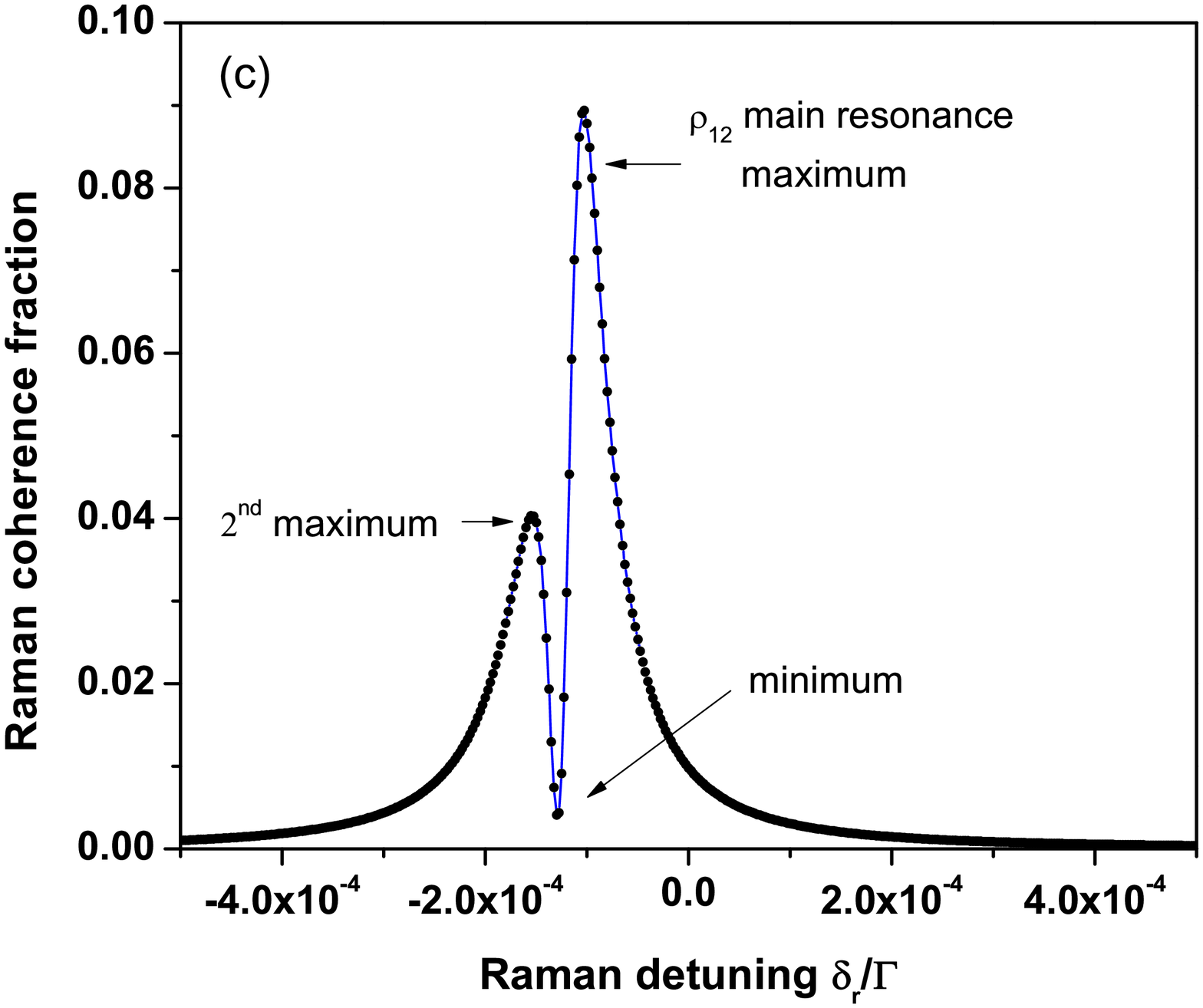}}
\caption{(Color online) (a) Steady-state lineshapes of the $|\rho_{12}|^{2}$ square modulus  versus $\delta_{r}$ using Eq.~(\ref{Raman-coherence}) for $\Gamma_{31}=\Gamma_{32}=\Gamma/2$, $\Omega_{1}=5\times10^{-2}\Gamma$, $\Omega_{2}=5\times10^{-3}\Gamma$, and $\gamma_{c}=0$. In (a) $\Delta_{0}=0$; in (b) $\Delta_{0}=5\Gamma$; in (c) $\Delta_{0}=20\Gamma$. Solid dots ($\bullet$) are from the numerical integration of Eq.~(\ref{set-Bloch}).}
\label{Raman-coherence-spectra}
\end{figure}
\begin{figure*}[!!t]
\resizebox{8.5cm}{!}{\includegraphics{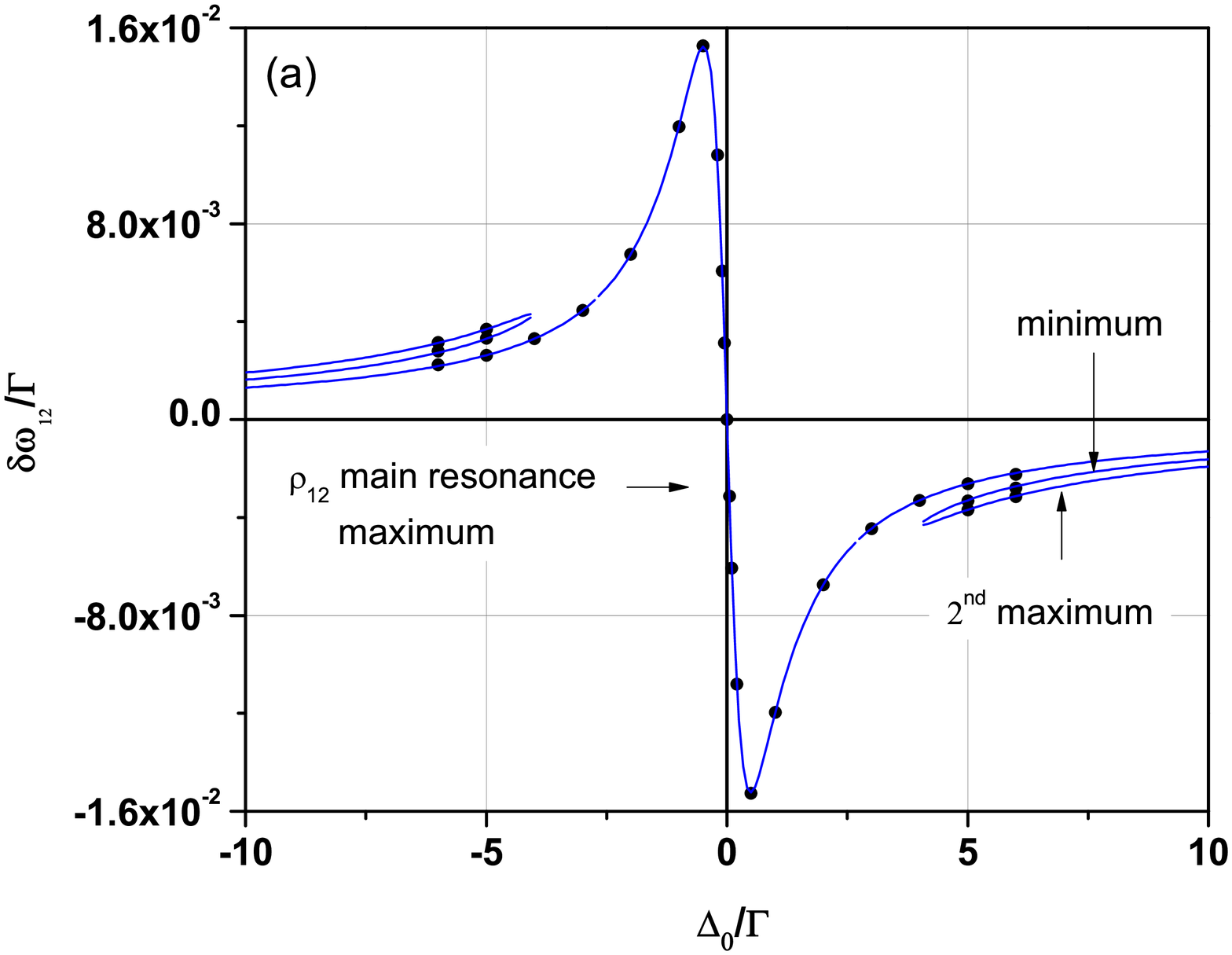}}\resizebox{8.5cm}{!}{\includegraphics{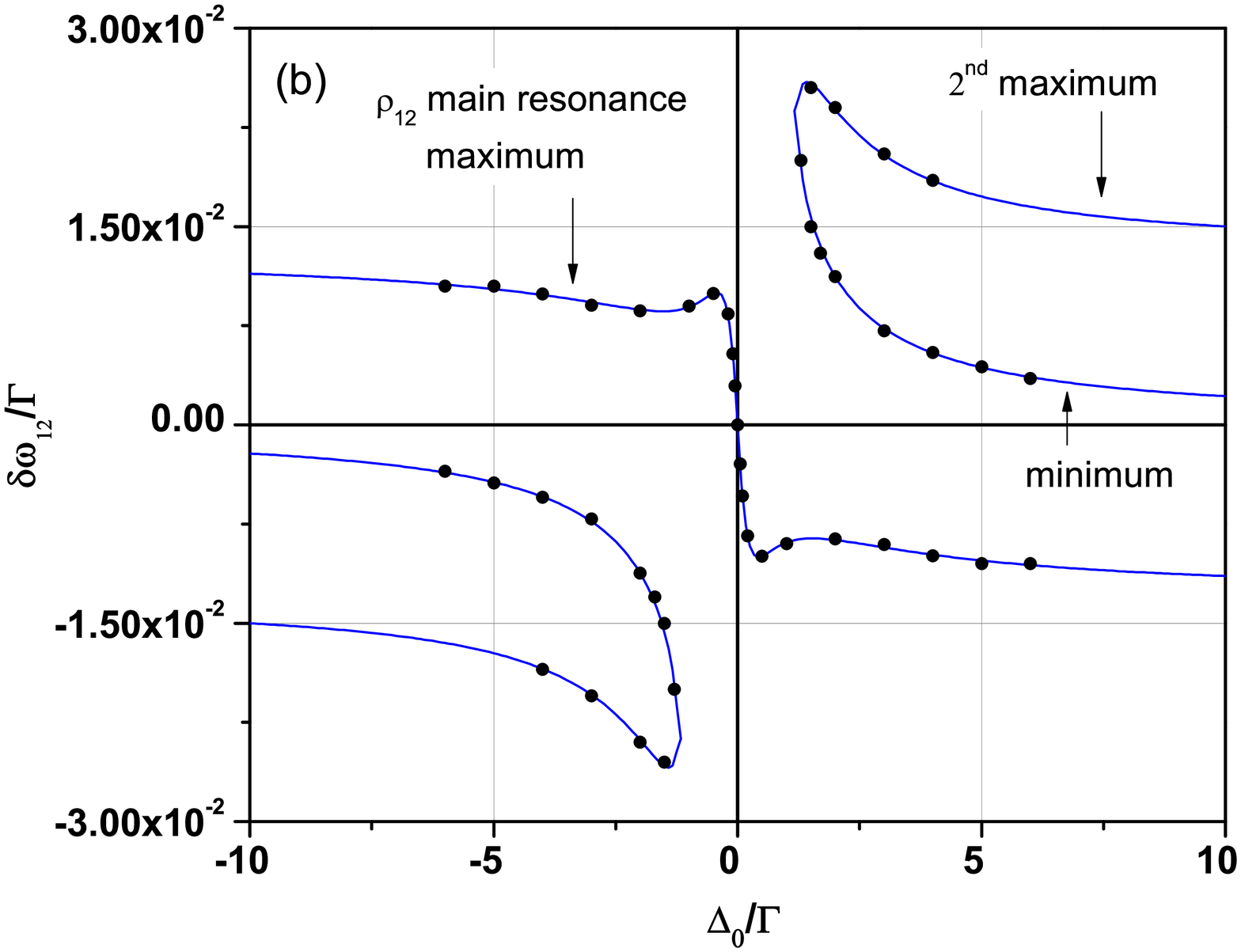}}
\caption{(Color online) (a) Frequency shift of the Raman coherence resonances observed on the $|\rho_{12}|^{2}$ as derived from
Eq.~(\ref{Raman-shift-coherence}) (solid blue line) versus the optical detuning $\Delta_{0}$. In (a) we have a symmetric radiative configuration with $\Gamma_{31}=\Gamma_{32}=\Gamma/2$ where $\Omega_{1}=5\times10^{-2}\Gamma$, $\Omega_{2}=5\times10^{-3}\Gamma$ and $\gamma_{c}=0$. In (b) we have an asymmetric radiative configuration with $\Gamma_{31}=\Gamma-\Gamma_{32}$, $\Gamma_{32}=10^{-5}\Gamma$ where $\Omega_{1}=5\times10^{-2}\Gamma$, $\Omega_{2}=5\times10^{-4}\Gamma$ and $\gamma_{c}=5~10^{-4}\Gamma$.
Solid dots ($\bullet$) are from the numerical integration of Eq.~(\ref{set-Bloch}).}
\label{Raman-coherence-shift}
\end{figure*}

\subsection{The Raman coherence resonance}

We are now focusing on the steady state Raman coherence resonance $\rho_{12}=Re\{\rho_{12}\}+i~Im\{\rho_{12}\}$ given by
\begin{eqnarray}
\begin{split}
Re\{\rho_{12}\}&=\Gamma_{12}S^{\Lambda}\frac{(\Delta_{1}-\Delta_{2})\overline{\Delta}-\gamma_{eff}}
{(\Delta_{1}-\Delta_{2})[\Delta_{1}-\Delta_{2}-\Delta_{f}]+\Gamma_{eff}^{2}},\\
Im\{\rho_{12}\}&=\Gamma_{12} S^{\Lambda}\frac{\left(\Delta_{1}-\Delta_{2}\right)+\overline{\Delta}\gamma_{c}}
{(\Delta_{1}-\Delta_{2})[\Delta_{1}-\Delta_{2}-\Delta_{f}]+\Gamma_{eff}^{2}},
\end{split}
\label{Raman-coherence}
\end{eqnarray}
with
\begin{equation}
\Gamma_{12}=\frac{\Omega_{2}^{2}\Gamma_{31}+\Omega_{1}^{2}\Gamma_{32}}{2\Omega_{1}\Omega_{2}}.
\end{equation}
When the dipole transition is allowed, the Raman coherence resonance can be detected in several manners. If we deal with alkaline atoms such as Cs or Rb, the hyperfine Raman coherence might be detected as a microwave emission proportional to $|\rho_{12}|^{2}$, inserting the atomic medium into a micro-wave cavity~\cite{Vanier:1998}. Fig.~\ref{Raman-coherence-spectra} shows $|\rho_{12}|^{2}$ versus the Raman detuning condition $\delta_{r}$ when $\Gamma_{31}=\Gamma_{32}=\Gamma/2$ at different values of $\Delta_{0}$.  The dispersive behavior of $Re\{\rho_{12}\}$ leads to a second resonant peak, appearing  when $\Delta_{0}>3\Gamma$ as seen in Fig.~\ref{Raman-coherence-spectra}(b) and Fig.~\ref{Raman-coherence-spectra}(c).

\subsection{Approximated frequency-shifts of the Raman coherence resonance}

We derive here an accurate expression for the frequency-shifted resonance of $|\rho_{12}|^{2}$ when $\Delta_{1}\sim\Delta_{2}\sim\Delta_{0}$. A cubic equation is derived from the analytical differentiation of $|\rho_{12}|^{2}$ of Eq.~(\ref{Raman-coherence}) with respect to the Raman detuning $\delta_r$. Using Cardan's solutions, the three roots are written  as:
\begin{eqnarray}
\begin{split}
&\delta\omega_{12}(\Delta_{0})\sim\\
&-\frac{b}{3a}+\sqrt{-\frac{p}{3}}~Cos\left[\frac{1}{3}Cos^{-1}\left[-\frac{3q}{p}\sqrt{-\frac{3}{p^{3}}}\right]+2k\frac{\pi}{3}\right].
\end{split}\label{Raman-shift-coherence}
\end{eqnarray}
with $k=0,1,2$, the $p$ and $q$ given by
\begin{eqnarray}
\begin{split}
p&=\frac{3ac-b^{2}}{3a^{2}}\\
q&=\frac{2b^{3}-9abc+27a^{2}d}{27a^{3}}
\end{split}\label{}
\end{eqnarray}
and
\begin{eqnarray}
\begin{split}
a=&-1-\overline{\Delta}^{2},\\
b=&3\overline{\Delta}\left(\gamma_{eff}-\gamma_{c}\right),\\
c=&-2\left(\gamma_{c}^{2}\overline{\Delta}^{2}+\gamma_{eff}^{2}\right)+\Gamma_{eff}^{2}\left(1+\overline{\Delta}^{2}\right)\\
&+\Delta_{f}\overline{\Delta}\left(\gamma_{c}+\gamma_{eff}\right),\\
d=&\Gamma_{eff}^{2}\overline{\Delta}\left(\gamma_{c}+\gamma_{eff}\right)+\Delta_{f}\left(\gamma_{c}^{2}\overline{\Delta}^{2}+\gamma_{eff}^{2}\right).
\end{split}\label{}
\end{eqnarray}
\indent Eq.~(\ref{Raman-shift-coherence}) allows us to obtain the frequency shift as a function of the common mode optical detuning $\Delta_{0}$ plotted in Fig.~\ref{Raman-coherence-shift}. However an estimate of that shift is obtained looking only at the real part of the coherence solution which mainly describes the lineshape emission. From the square modulus of the real part $|Re\{\rho_{12}\}|^{2}$, simple cubic solutions for the Raman coherence frequency-shift can be derived as:
\begin{eqnarray}
\begin{split}
&\delta\omega_{12}(\Delta_{0})\sim\\
&\frac{\gamma_{eff}}{\overline{\Delta}}\left(1\mp\sqrt{1+\frac{\overline{\Delta}\Gamma_{eff}}{\gamma_{eff}}\left(\frac{\overline{\Delta}\Gamma_{eff}}{\gamma_{eff}}-\frac{\Delta_{f}}{\Gamma_{eff}}\right)}\right),
\end{split}\label{}
\end{eqnarray}
or
\begin{equation}
\delta\omega_{12}(\Delta_{0})\sim\frac{\gamma_{eff}}{\overline{\Delta}}.
\label{}
\end{equation}
When $\overline{\Delta}\Gamma_{eff}/\gamma_{eff}\ll1$ and $\Delta_{f}/\Gamma_{eff}\ll1$, the Raman coherence frequency-shift expression corresponding to the maximum emission becomes:
\begin{equation}
\delta\omega_{12}(\Delta_{0})\sim\frac{\Delta_{f}}{2}.
\label{}
\end{equation}
For that case, we recover the usual dispersive shape related to the light-shift $\Delta_{LS}$ affecting clock states.
\indent The frequency shift versus the common mode optical detuning $\Delta_{0}$ derived from Eq.~(\ref{Raman-shift-coherence}) is shown in Fig.~\ref{Raman-coherence-shift}.  for the case of  a symmetric radiative configuration with $\Gamma_{31}=\Gamma_{32}$ while in (b) for $\Gamma_{31}\gg\Gamma_{32}$. The central dispersive feature, related to the Raman shift expression $\Delta_{f}$, corresponds to the maximum of the coherent emission. Other  branches of the shift correspond to the extrema of the second resonance appearing for $\Delta_{0}>3\Gamma$ as from the lineshape simulation of Fig.~\ref{Raman-coherence-spectra}(b). A direct comparison of Fig.~\ref{Raman-coherence-shift}(b) with frequency shifts reported in Fig.~\ref{shift-population} with similar conditions, yield to Raman coherence shifts larger by more than an order of magnitude than population frequency shifts.

\section{Dark Resonance fringes}

\subsection{Pulsed regime lineshape}
\begin{figure}[!!t]
\centering
\resizebox{9.0cm}{!}{\includegraphics[angle=-90]{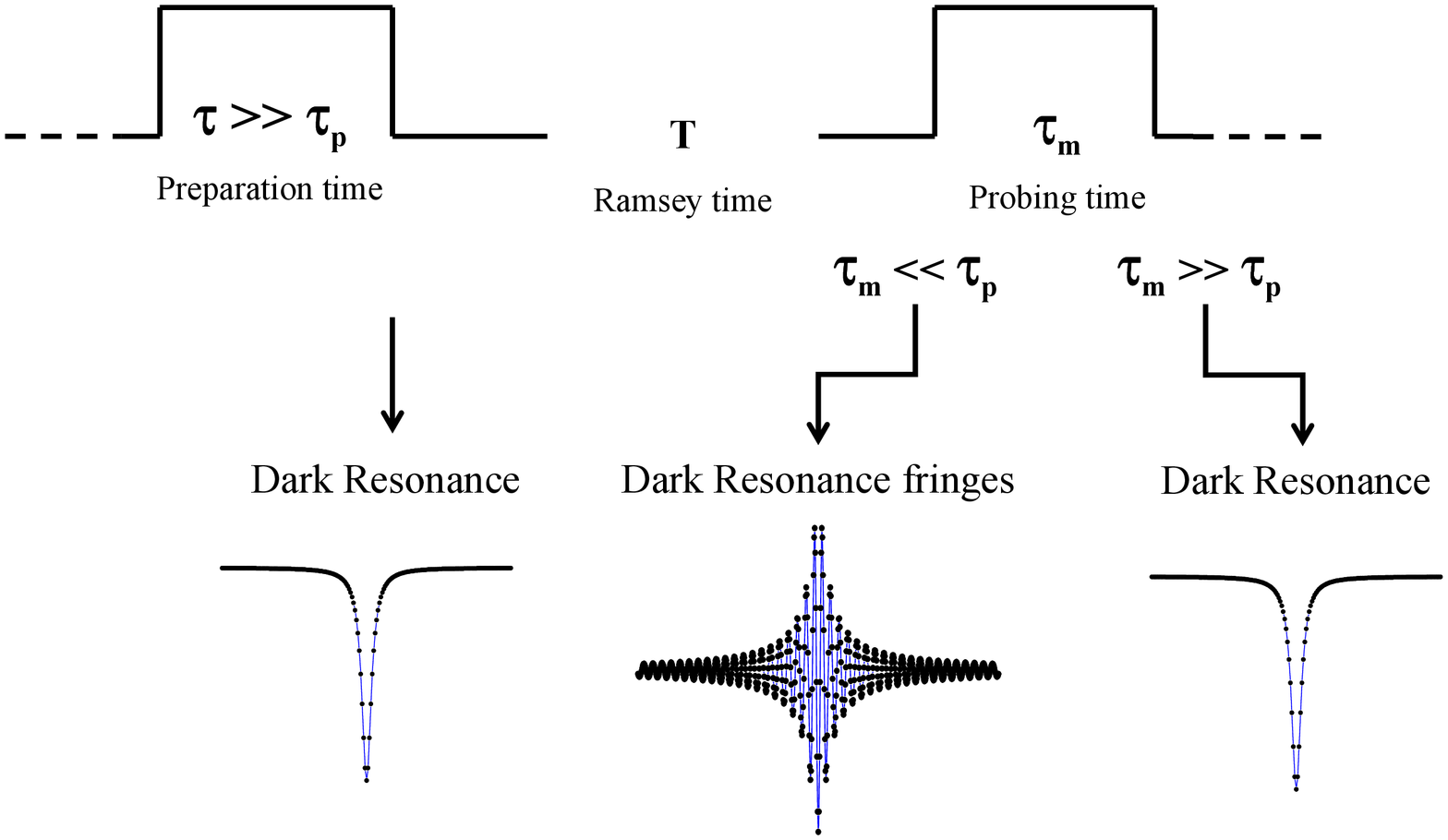}}
\caption{(Color online) Pulsed Dark Resonance detection scheme to perform high resolution spectroscopy of three-level $\Lambda$ systems. $T$ is the Ramsey time when both laser light fields $\Omega_{1},\Omega_{2}$ are switched off. The first pulse is long enough $\tau\gg\tau_{p}(\Delta_{0})$ to reach the stationary regime. During the second pulse, the probing time can be short $\tau_{m}\ll\tau_{p}(\Delta_{0})$ to observe Dark Resonance fringes or long $\tau_{m}\gg\tau_{p}(\Delta_{0})$ to recover a cw Dark Resonance as a new preparation stage for the next pulse.}
\label{interrogation-scheme}
\end{figure}

The clock operation may be based on a pulsed Raman-Ramsey scheme, illustrated in Fig.~\ref{interrogation-scheme}, with beating
oscillations observed whichever variable is monitored. This detection approach
originally introduced in~\cite{Thomas:1982}, was refined in~\cite{Zanon:2005-PRL,Zanon:2005-IEEE} and
discussed in refs~\cite{Zanon-Willette:2006,Yoon:2007,Yudin:2010}. It
allows to reach a higher precision in the clock frequency measurement, as typical of the
Ramsey fringes. The present work focuses on the laser pulse scheme where the first
pulse is long enough to allow the atomic or molecular preparation into the dark state superposition and the second delayed short pulse probing that superposition. Because the $\tau_{p}$ pumping time  of Eq.~\eqref{time-population} is required to reach the steady-state atomic or molecular preparation into the dark state, the length of the first pulse should satisfy $\tau\gg\tau_{p}(\Delta_{0})\gg\Gamma_{31}^{-1},\Gamma_{32}^{-1}$.
From a mathematical point of view the $\tau\mapsto\infty$ steady state solution of the
three-level optical Bloch equations may thus be used as initial condition for determining the
evolution at later times. At time $\tau$ the laser fields are switched off in order to allow
for a free evolution over the time \emph{T}. Finally a readout operation is performed by applying a short pulse whose time duration $\tau_{m}$ is limited by
$\Gamma_{31}^{-1},\Gamma_{32}^{-1}<\tau_{m}\ll\tau_{p}(\Delta_{0})$. In this adiabatic regime where
$d\rho_{33}(t)/dt=d\rho_{13}(t)/dt=d\rho_{23}(t)/dt\equiv0$, the short probe pulse operation is well described
using the $\tau_{m}\mapsto0$ limit. For a readout pulse as long as the preparation pulse, all atoms or molecules are repumped into a new dark state erasing interference fringes.
Interferences fringes are detectable on all observables as a function of the $T$ time delay,  with very short readout pulses required for optical coherences and the excited state population fraction. Instead longer probing times are required for monitoring fringes on lower state populations due to slow time evolution of the clock states and the Raman coherence.\\
\begin{figure}[!t]
\resizebox{9.0cm}{!}{\includegraphics[angle=0]{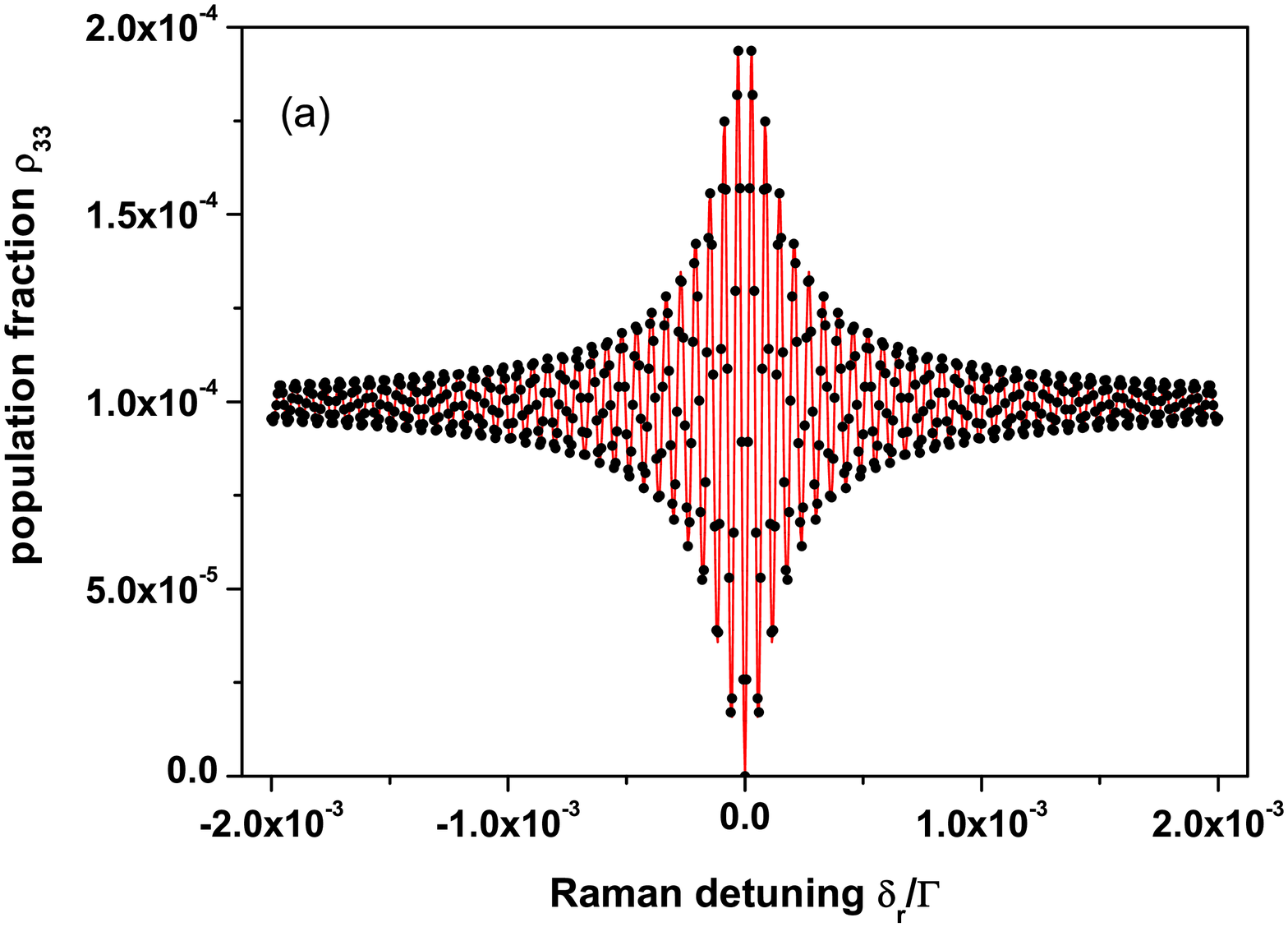}}
\resizebox{9.0cm}{!}{\includegraphics[angle=0]{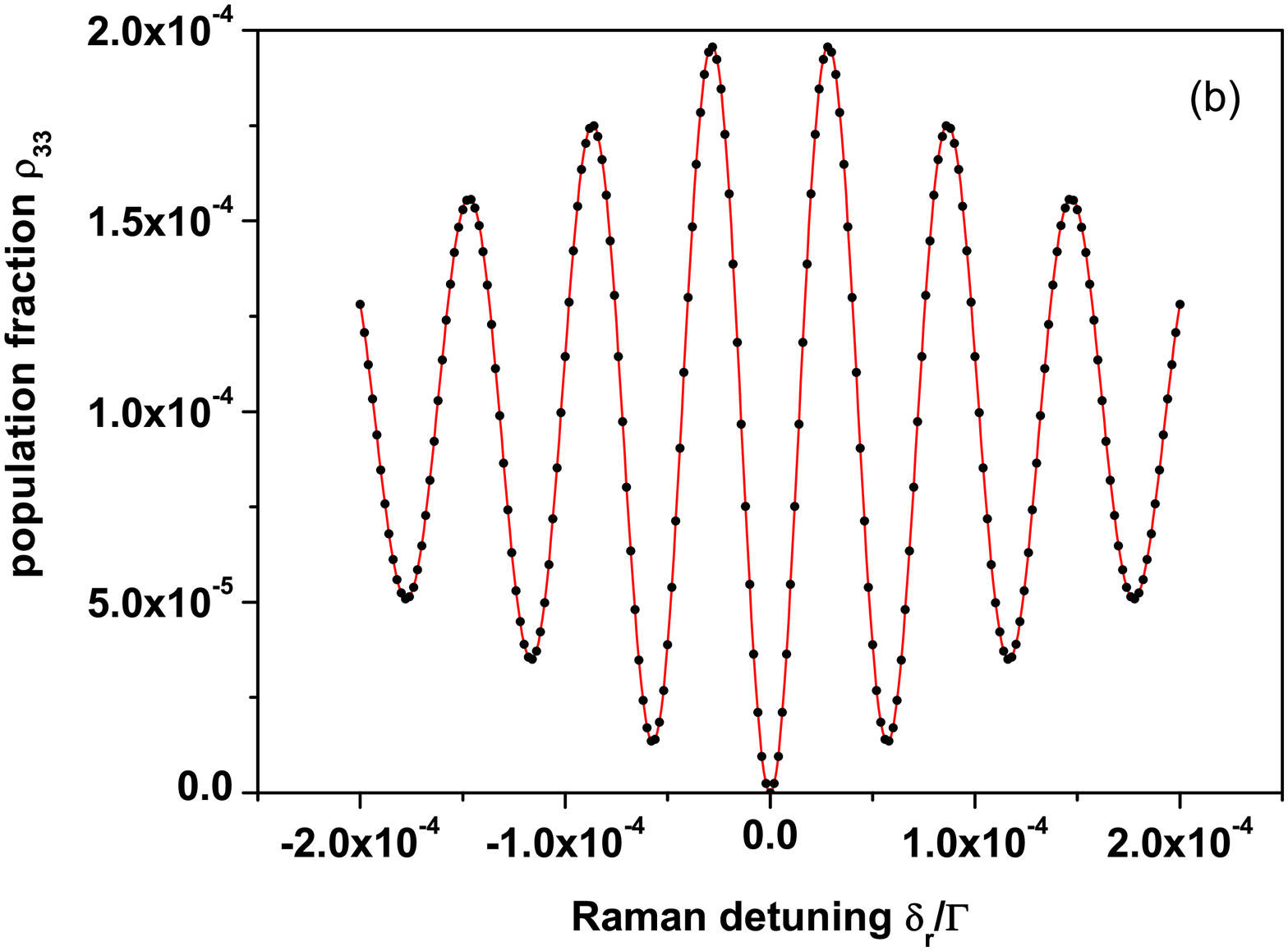}}
\caption{(Color online) (a) Dark Resonance fringes in the weak field limit. (b) Frequency span of the lineshape. Lines from Eq.~(\ref{Pulsed-Dark-Resonance}) and dots $(\bullet)$ from Bloch's Eqs.~(\ref{set-Bloch}). Common parameters: $\Gamma_{31}=\Gamma_{32}=\Gamma/2$, $\Delta_{0}=0$, $\gamma_{c}=0$ and $T=10\tau_{p}(0)$. In (a) and (b) Rabi frequencies $\Omega_{1}=\Omega_{2}=0.005\Gamma$, free evolution time  $T=10~\tau_{p}(0)$, probe time $\tau_{m}\sim 0.004~\tau_{p}(0)$. Very good agreement between  Eq.~(\ref{set-Bloch}) and  Eq.~(\ref{Pulsed-Dark-Resonance}) results.}
\label{pulsed-Dark-resonance}
\end{figure}
\begin{figure}[!!t]
\resizebox{8.5cm}{!}{\includegraphics[angle=0]{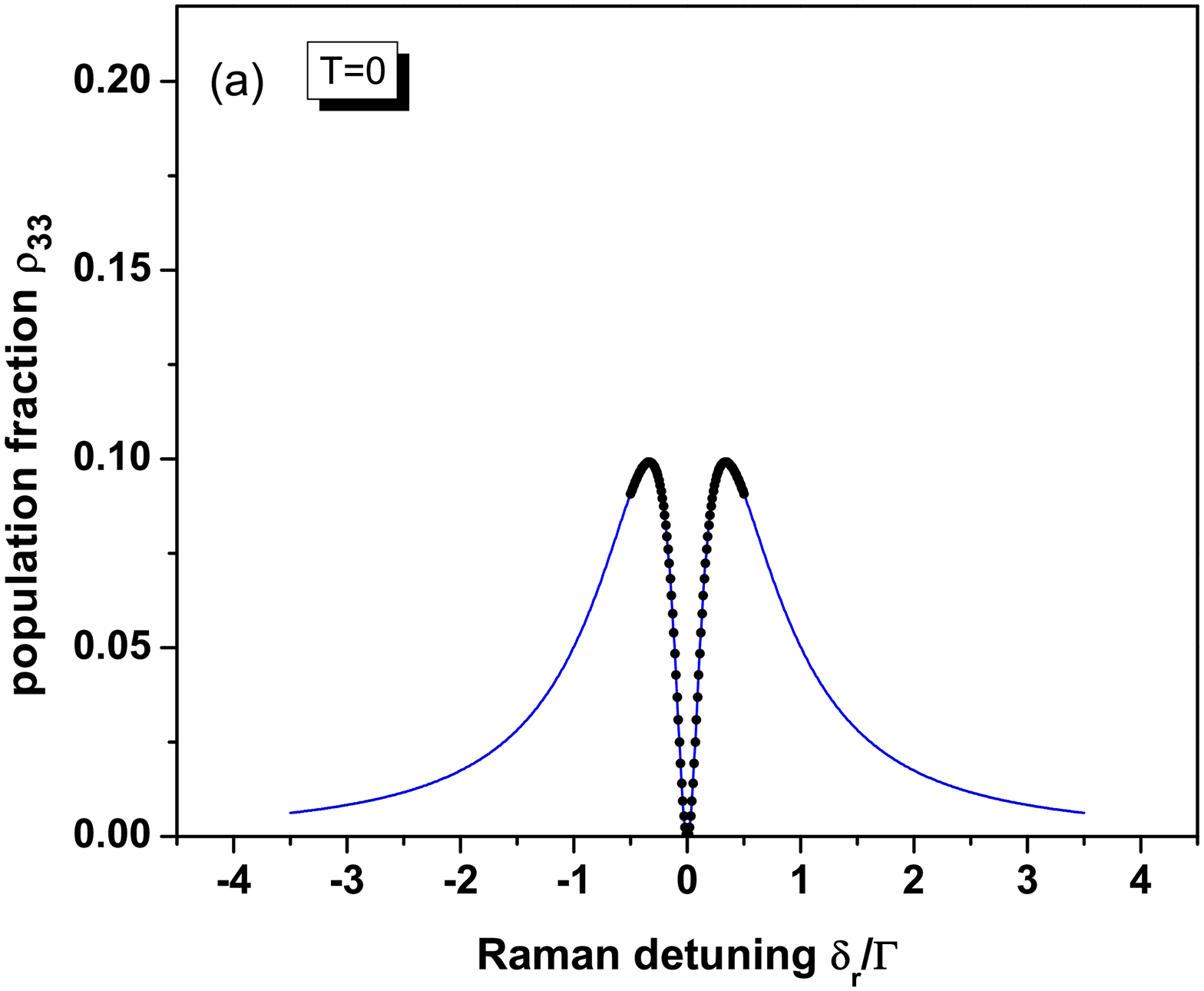}}
\resizebox{8.5cm}{!}{\includegraphics[angle=0]{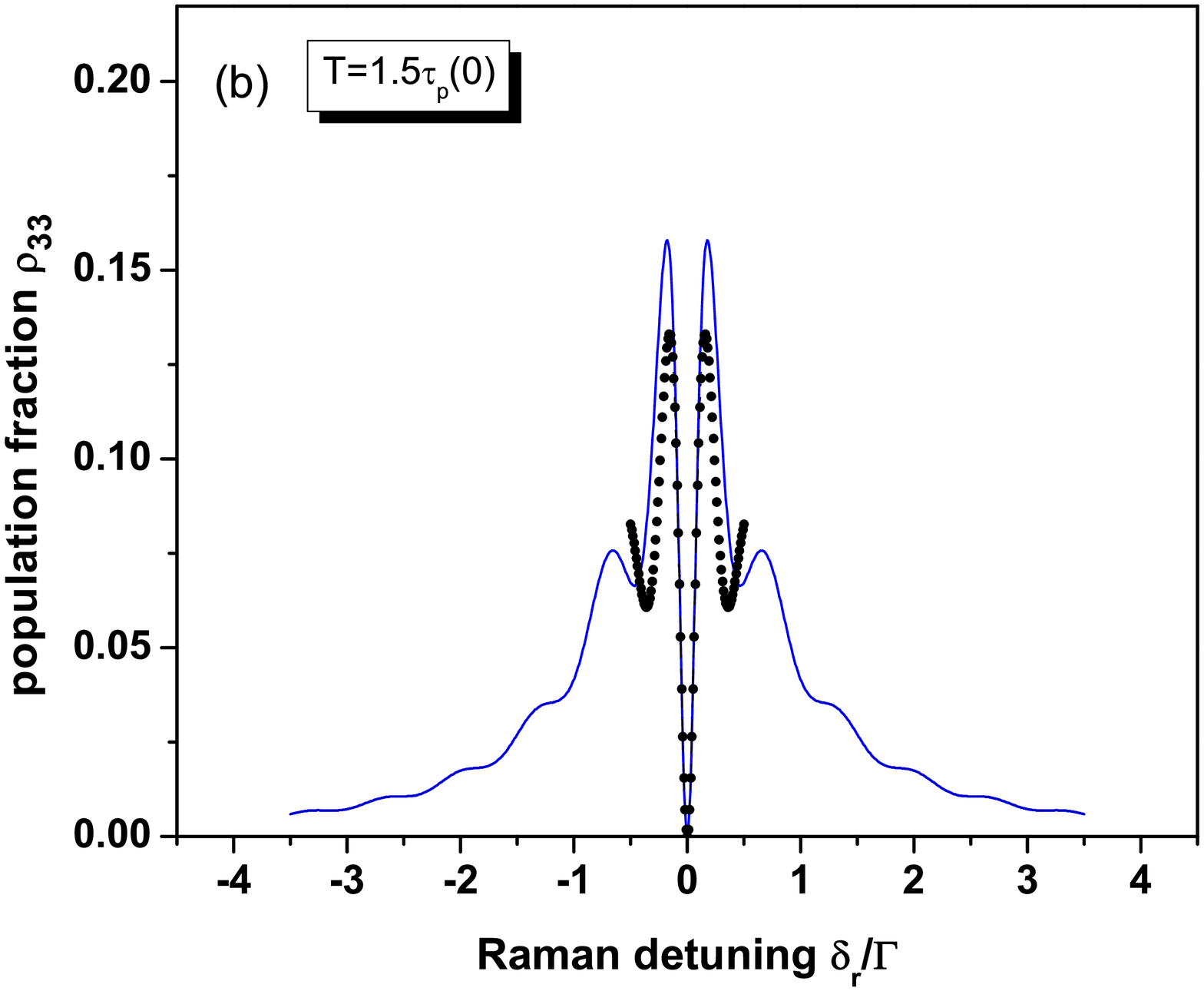}}
\resizebox{8.5cm}{!}{\includegraphics[angle=0]{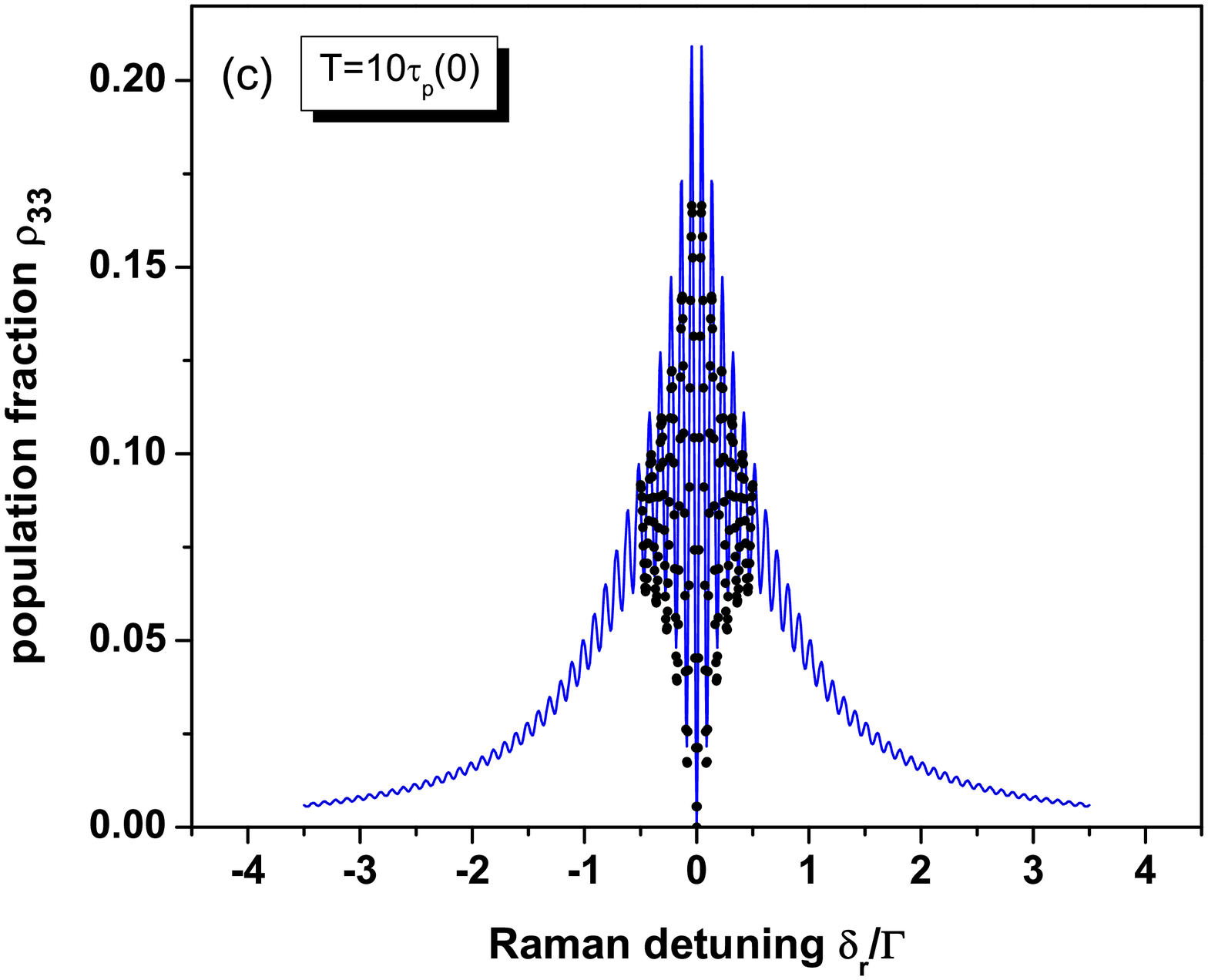}}
\caption{(Color online) DR lineshapes in strong laser fields computed from Eq.~(\ref{Pulsed-Dark-Resonance}) versus Raman detuning for different Ramsey time $T$. Parameters $\Omega_{1}=\Omega_{2}=0.2\Gamma$, $\Gamma_{31}=\Gamma_{32}=\Gamma/2$, $\Delta_{0}=0$ and $\gamma_{c}=0$. Solid dots $(\bullet)$ from Bloch's Eqs.~(\ref{set-Bloch}) with $\tau=10~\tau_{p}(0)$.  In (b) and (c) $\tau_{m}=0.75~\tau_{p}(0)$.}
\label{amplitude-fringes}
\end{figure}
\begin{figure*}[!!t]
\centering\resizebox{18cm}{!}{\includegraphics[angle=0]{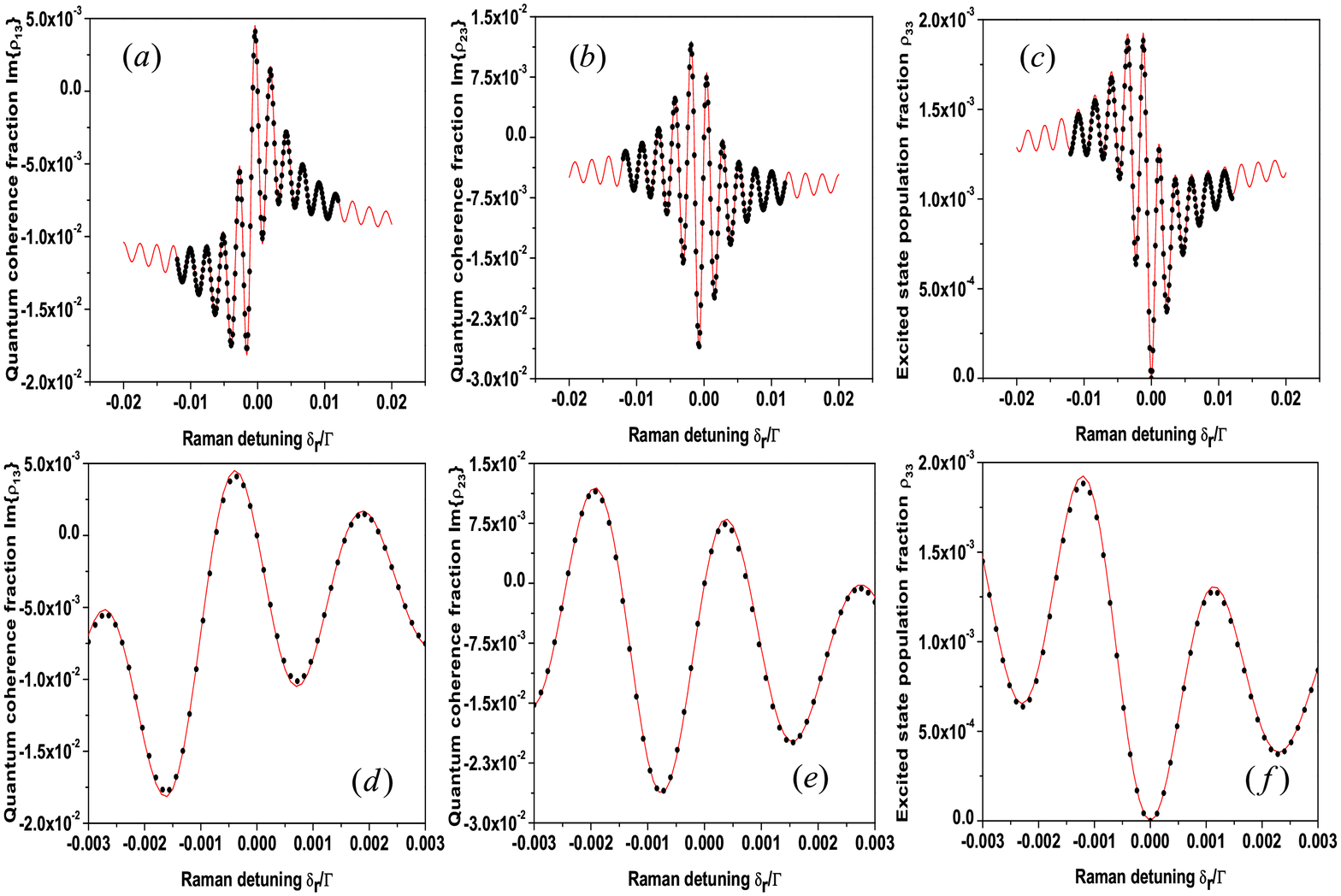}}
\caption{(Color online) Asymmetric fringes observed on the imaginary parts of optical coherences, $Im\{\rho_{13}\}(T)$ in (a) and $Im\{\rho_{23}\}(T)$ in (b), and on the excited population $\rho_{33}$, in (c), versus Raman frequency detuning. In (d), (e), and (f) expanded views of  the central resonance fringe for each observable. Parameters $\Gamma_{31}=0.8\Gamma$, $\Gamma_{32}=0.2\Gamma$, $\Delta_{0}=\Gamma$, $\gamma_{c}=0$, $\Omega_{1}=5.10^{-2}\Gamma$, $\Omega_{2}=2.5\times10^{-2}\Gamma$, free evolution time $T=10~\tau_{p}(0)$, $\tau>10\tau_{p}(\Gamma)$, $\tau_{m}\sim 5.5\times10^{-3}\tau_{p}(\Gamma)$. Solid lines from the analytical forms of Eqs.~(\ref{imaginary-part-13}),~(\ref{imaginary-part-23}), and \eqref{Pulsed-Dark-Resonance}, respectively, in good agreement with the solid dots ($\bullet$) results from Bloch's Eqs.~(\ref{set-Bloch}).}
\label{fringes}
\end{figure*}
\indent An analytical expression of the population fraction in the excited state can be established from Eq.~(\ref{set-Bloch}) in the
asymptotic limit of a long preparation pulse and a vanishing read
out pulse, i.e. at the end of the free evolution time. In this regime, a pulsed DR expression
$\rho_{33}(T)$ and related transmission parts of the $Im\{\rho_{13}\}(T),Im\{\rho_{23}\}(T)$ optical
coherences
can be rewritten in the exact $\alpha_{ij}\left[1+\beta_{ij}|\rho_{12}|e^{-\gamma_{c}T}\cos\left(\delta_{r}
T-\Phi_{ij}\right)\right]$ form:
\begin{eqnarray}
\begin{split}
\rho_{33}(T)&=\alpha_{33}\times\left(1+\beta_{33}|\rho_{12}|e^{-\gamma_{c}T}\cos\left(\delta_{r}
T-\Phi_{33}\right)\right),\\\\
\alpha_{33}&=\frac{\Omega_{2}^{2}(\Delta
n+1)/\Gamma\widetilde{\gamma_{2}}-\Omega_{1}^{2}(\Delta
n-1)/\Gamma\widetilde{\gamma_{1}}}
{1+\frac{3}{\Gamma}(\Omega_{1}^{2}/\widetilde{\gamma_1}
+\Omega_{2}^{2}/\widetilde{\gamma_2})},\\
\beta_{33}&=\frac{\sqrt{\mu_{\gamma33}^{2}+\mu_{\Delta33}^{2}}}{\alpha_{33}}.
\end{split}\label{Pulsed-Dark-Resonance}
\end{eqnarray}
\begin{eqnarray}
\begin{split}
Im\{\rho_{13}\}(T)=&\alpha_{13}\times\left(1+\beta_{13}|\rho_{12}|e^{-\gamma_{c}T}\cos\left(\delta_{r}
T-\Phi_{13}\right)\right),\\\\
\alpha_{13}=&\frac{\Omega_{1}}{2\widetilde{\gamma_1}}\left(3\alpha_{33}+(\Delta n-1)\right),\\
\beta_{13}=&\frac{\sqrt{\mu_{\gamma13}^{2}+\mu_{\Delta13}^{2}}}{\alpha_{13}}.
\end{split}\label{imaginary-part-13}
\end{eqnarray}
\begin{eqnarray}
\begin{split}
Im\{\rho_{23}\}(T)&=\alpha_{23}\times\left(1+\beta_{23}|\rho_{12}|e^{-\gamma_{c}T}\cos\left(\delta_{r}
T-\Phi_{23}\right)\right),\\\\
\alpha_{23}=&\frac{\Omega_{2}}{2\widetilde{\gamma_2}}\left(3\alpha_{33}-(\Delta n+1)\right),\\
\beta_{23}=&\frac{\sqrt{\mu_{\gamma23}^{2}+\mu_{\Delta23}^{2}}}{\alpha_{23}}.
\end{split}\label{imaginary-part-23}
\end{eqnarray}
where $\Delta n=\rho_{22}-\rho_{11}$ is the steady-state clock state population difference obtained from
Eq.~\eqref{populations} and $|\rho_{12}|$ given by Eq.~\eqref{Raman-coherence}.
The following quantities were introduced in the above expressions:
\begin{eqnarray}
\begin{split}
&\mu_{\gamma33}=\frac{2\Omega_{1}\Omega_{2}}{\Gamma}\left(\frac{(\widetilde{\gamma_{1}}+\widetilde{\gamma_{2}})/
\widetilde{\gamma_{1}}\widetilde{\gamma_{2}}}{1+\frac{3}{\Gamma}(\Omega_{1}^{2}/\widetilde{\gamma_1}
+\Omega_{2}^{2}/\widetilde{\gamma_2})}\right),\\
&\mu_{\Delta33}=\frac{2\Omega_{1}\Omega_{2}}{\Gamma}\left(\frac{\Delta_{1}
/\widetilde{\gamma_{1}}\gamma_{1}-\Delta_{2}/\widetilde{\gamma_{2}}\gamma_{2}}{1+\frac{3}{\Gamma}(\Omega_{1}^{2}/\widetilde{\gamma_1}
+\Omega_{2}^{2}/\widetilde{\gamma_2})}\right),
\end{split}
\end{eqnarray}
\begin{eqnarray}
\begin{split}
&\mu_{\gamma13}=\frac{3}{2}\frac{\Omega_{1}}{\widetilde{\gamma_1}}\mu_{\gamma33}
-\frac{\Omega_{2}}{\widetilde{\gamma_1}},\\
&\mu_{\Delta13}=\frac{3}{2}\frac{\Omega_{1}}{\widetilde{\gamma_1}}\mu_{\Delta33}
-\Omega_{2}\frac{\Delta_{1}}{\gamma_1\widetilde{\gamma_1}},
\end{split}
\end{eqnarray}
\begin{eqnarray}
\begin{split}
&\mu_{\gamma23}=\frac{3}{2}\frac{\Omega_{2}}{\widetilde{\gamma_2}}\mu_{\gamma33}
-\frac{\Omega_{1}}{\widetilde{\gamma_2}},\\
&\mu_{\Delta23}=\frac{3}{2}\frac{\Omega_{2}}{\widetilde{\gamma_2}}\mu_{\Delta33}
+\Omega_{1}\frac{\Delta_{2}}{\gamma_2\widetilde{\gamma_2}}.
\end{split}
\end{eqnarray}
In the non asymptotic limit where
$\tau\sim\tau_{m}\sim\tau_{p}$, stationary solutions of
$|\rho_{12}|$ and $\Delta n$ have to be replaced by their transient
expressions $|\rho_{12}(\tau,\tau_{m})|$ and $\Delta
n(\tau,\tau_{m})$. When, $T\mapsto0$, the lineshape expression from Eq.~(\ref{Pulsed-Dark-Resonance}) is
formally equivalent to Eq.~(\ref{fluorescence-solution}).
\indent A pulsed DR lineshape originated is plotted in
Fig.~\ref{pulsed-Dark-resonance}(a) for small values of Rabi frequencies with a magnified span on the
central fringe in Fig.~\ref{pulsed-Dark-resonance}(b). In these plots, the
agreement between Eq.~(\ref{Pulsed-Dark-Resonance}) (solid red
line) and the Bloch's equations (dots $\bullet$) is very accurate.
We have also studied the role of the Ramsey time $T$ on the DR lineshape and amplitude. Fig.~\ref{amplitude-fringes}(a)-(c) shows the signal amplitude versus the Raman detuning for different values of $T$. The fringe amplitude is always twice the amplitude in the cw regime when $2T>\Gamma_{eff}^{-1}$ and $\gamma_{c}=0$. A careful comparison between Fig.~\ref{pulsed-Dark-resonance} and Fig.~\ref{amplitude-fringes}(b) and Fig.~\ref{amplitude-fringes}(c) show that for Rabi frequencies comparable to the $\Gamma_{31},\Gamma_{32}$ relaxation rates, i.e., when saturation becomes important, the limit $\tau_{m}\mapsto0$
used to obtain Eq.~(\ref{Pulsed-Dark-Resonance}) is no longer valid and the effect of the readout duration pulse length $\tau_{m}$ has to be included into the analysis leading to a slight reduction of the fringe amplitude.\\

\subsection{Stationary atomic/molecular Raman phase}
The fringe signals appearing in the atomic observables are produced by the $\cos(\delta_{r}
T-\Phi_{ij})$ terms appearing in the above expressions. Within a narrow region around the resonance, the lineshape is entirely described by a phase given by
\begin{equation}
\begin{split}
\Phi_{33}&=Arctan\left[\frac{\mu_{\gamma33}~Im\{\rho_{12}\}-\mu_{\Delta33}~Re\{\rho_{12}\}}
{\mu_{\gamma33}~Re\{\rho_{12}\}+\mu_{\Delta33}~Im\{\rho_{12}\}}\right], \\
\Phi_{13}&=Arctan\left[\frac{\mu_{\gamma13}~Im\{\rho_{12}\}-\mu_{\Delta13}~Re\{\rho_{12}\}}
{\mu_{\gamma13}~Re\{\rho_{12}\}+\mu_{\Delta13}~Im\{\rho_{12}\}}\right], \\
\Phi_{23}&=Arctan\left[\frac{\mu_{\gamma23}~Im\{\rho_{12}\}-\mu_{\Delta23}~Re\{\rho_{12}\}}
{\mu_{\gamma23}~Re\{\rho_{12}\}+\mu_{\Delta23}~Im\{\rho_{12}\}}\right].
\end{split} \label{phase-shift}
\end{equation}
\begin{figure}[!t]
\centering\resizebox{9.0cm}{!}{\includegraphics[angle=0]{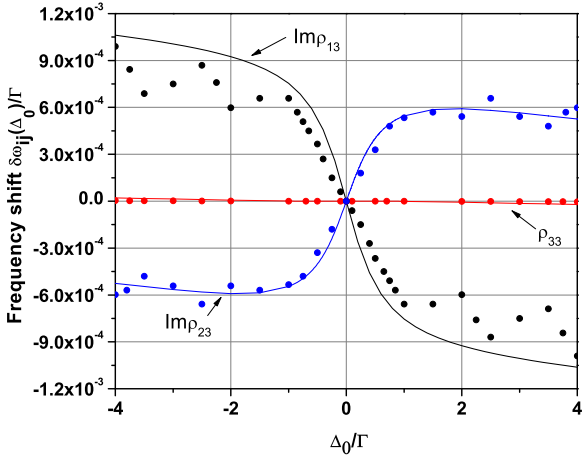}}
\caption{(Color online) Exact numerical tracking of fringe frequency-shifts $\delta\omega^{\rm fr}_{13}$, $\delta\omega^{\rm fr}_{23}$ and $\delta\omega^{\rm fr}_{33}$ versus $\Delta_0$ derived from Eqs.~(\ref{Pulsed-Dark-Resonance}), (\ref{imaginary-part-13}) and
 (\ref{imaginary-part-23}). Parameters $\Gamma_{31}=0.8\Gamma$, $\Gamma_{32}=0.2\Gamma$, $\gamma_{c}=5\times10^{-4}\Gamma$,  $\Omega_{1}=5\times10^{-2}\Gamma$, $\Omega_{2}=3.7515\times10^{-2}\Gamma$, $T=10~\tau_{p}(0)$. Solid dots ($\bullet$) from Bloch's equations~(\ref{set-Bloch}) where we use $\tau>10\tau_{p}(\Gamma)$, $\tau_{m}\sim 5.5\times10^{-3}\tau_{p}(\Gamma)$. Discrepancy between solid lines and dots are here due to the non vanishing $\tau_{m}$ readout time required with a numerical integration of Bloch's equations.}
\label{fringes-shifts}
\end{figure}
\indent In the adiabatic regime, for a first pulse producing a steady state and a second pulse with a short duration, the $|\rho_{12}|$ steady state solution of Eq.~\eqref{Raman-coherence} leads to cw expressions of $\Phi_{ij}$ $(i,j=1,2,3)$ where all time dependances are absent. For clock engineering, it is important to quantify the shift of the central fringe when the common mode optical detuning $\Delta_{1}\sim\Delta_{2}\sim\Delta_{0}$ is scanned around the $\delta_r=0$ resonant value. Within the limit of a long first pulse and a short second pulse, in the weak field regime and for a sufficiently large Ramsey time ($T\gg\tau_{p}(\Delta_{0}),\tau_{m}$), the central fringe frequency-shift $\delta\omega^{\rm fr}_{ij}$ should be connected to the $\Phi_{ij}$ phase accumulation. Time-dependent Raman frequency-shifts produced by short preparation pulses were examined in details in ref.~\cite{Hemmer:1989,Shahriar:1997} using the Bloch vector model, but only for the excited state population. A complex wave-function formalism was also proposed for short pulses in~\cite{Zanon-Willette:2006} and extended in ref.~\cite{Yoon:2007} to derive analytical time-dependent expressions of any frequency-shift affecting the central fringe related to each possible observable.\\
\indent Fig.~\ref{fringes}(a)-(c) reports the fringes appearing on the imaginary parts $Im\{\rho_{13}\}(T)$, $Im\{\rho_{23}\}(T)$ and on the excited state population $\rho_{33}(T)$ for an asymmetrical radiative configuration with $\Gamma_{31}=0.8\Gamma$ and $\Gamma_{32}=0.2\Gamma$, a detuned laser excitation ($\Delta_{0}=\Gamma$) with no Raman decoherence ($\gamma_{c}=0$). Plots from Fig.~\ref{fringes}(d)-(f) report a magnified span on the central fringe. The plots in Fig.~\ref{fringes}(a),(b) for the optical coherences show asymmetrical lineshapes with the central fringe being blue or red shifted from the exact Raman resonance, a behaviour different from that observed in the cw regime. For the excited state fraction plotted in Fig.~\ref{fringes}(c) under same conditions, the central fringe is not frequency shifted, as in the cw regime. These results point out, for the first time, that the imaginary parts of optical coherences, when individually probed, have a lineshape different from that observed on the excited state response due to asymmetric decay rates by spontaneous emission.\\
\begin{figure}[!t]
\centering\resizebox{9.0cm}{!}{\includegraphics[angle=0]{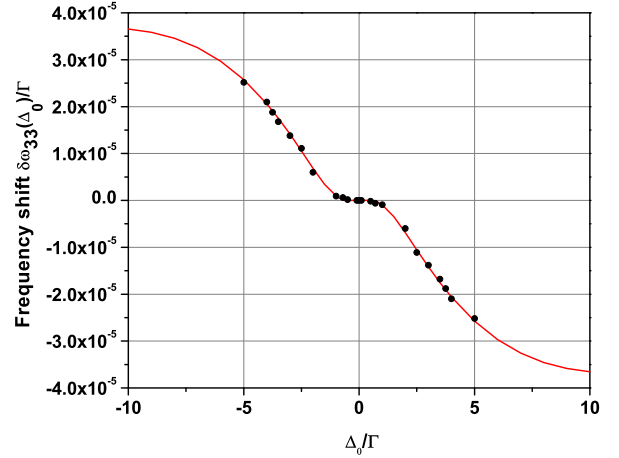}}
\caption{(Color online) Exact numerical tracking of the Dark Resonance fringe frequency-shift $\delta\omega^{\rm fr}_{33}$ versus $\Delta_0$ derived from Eqs.~(\ref{Pulsed-Dark-Resonance}). Solid dots ($\bullet$) are plotted from Bloch's Eqs.~(\ref{set-Bloch}) with parameters as in Fig.~\ref{fringes-shifts}.}
\label{excited-state-fringe-shift}
\end{figure}
\begin{figure}[!b]
\resizebox{9.0cm}{!}{\includegraphics[angle=0]{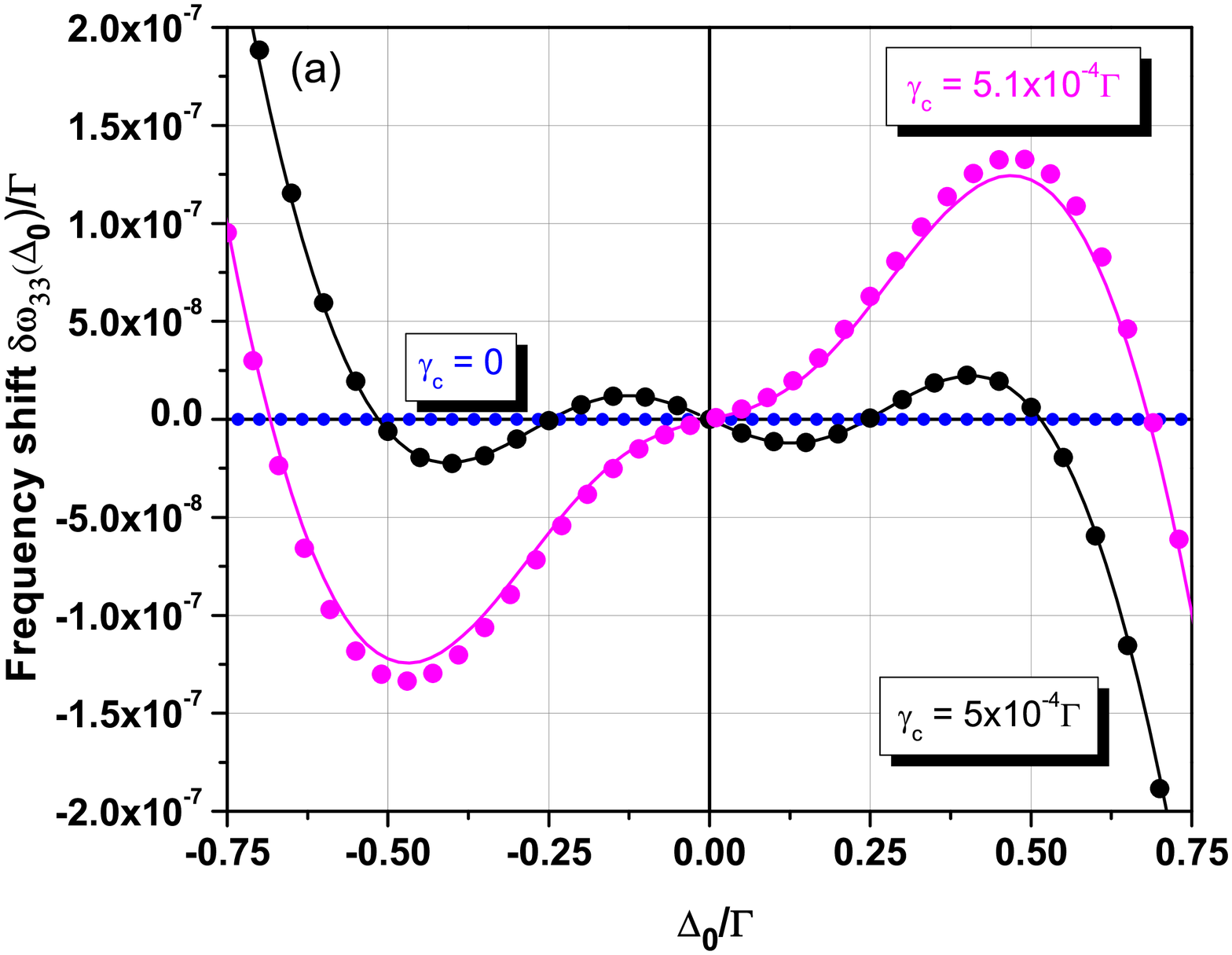}}
\resizebox{9.0cm}{!}{\includegraphics[angle=0]{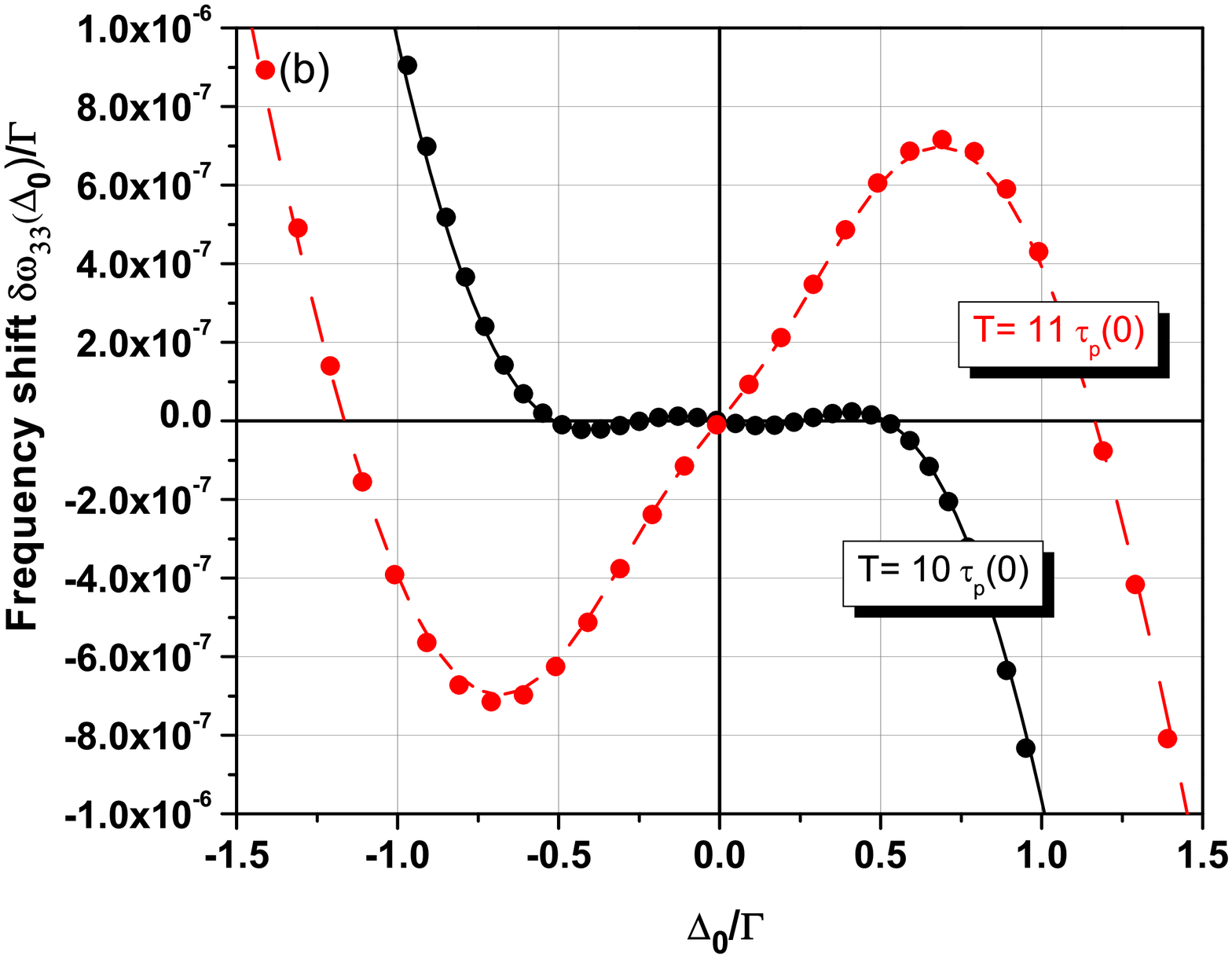}}
\caption{(Color online) (a) Expanded view of the fringe frequency-shift numerical tracking $\delta\omega^{\rm fr}_{33}$ from Fig.~\ref{excited-state-fringe-shift} for various values of the small decoherence $\gamma_{c}$ term. (b) Exact Numerical frequency-shift $\delta\omega^{\rm fr}_{33}(\Delta_{0})$ with different Ramsey time $T$ for the particular ratio $\Omega_{2}/\Omega_{1}\sim0.75$ where $\Omega_{1}=5\times10^{-2}\Gamma$ and $\gamma_{c}=5\times10^{-4}\Gamma$. Solid dots in (a) and (b) cases are plotted from the analytical first order expansion of the Raman shift using Eq.~(\ref{pulsed-fringe-frequency-shift}). Note the oscillating pattern with small amplitude.}
\label{expanded-fringe-shift}
\end{figure}
\indent This different behaviour is also confirmed by the plots of Fig.~\ref{fringes-shifts} focusing on Dark Resonance fringe frequency-shifts versus the common mode optical detuning $\Delta_{0}$ when assuming a small decoherence term $\gamma_{c}=5\times10^{-4}\Gamma$. Fig.~\ref{fringes-shifts} are numerical tracking of frequency-shifts from the Eq.~(\ref{Pulsed-Dark-Resonance}) analytical form  compared to the integration of Bloch's equations~(\ref{set-Bloch}) (solid dots) for a particular ratio between Rabi frequencies. Oscillations around the numerical track of frequency shifts are observed for large optical detunings due to the non vanishing $\tau_{m}$ readout time used with Eqs.~(\ref{set-Bloch}). Fig.~\ref{excited-state-fringe-shift} evidences a very weak slope near optical resonance; such dependence does not appear on the imaginary parts of the optical transmission.  In absence of  the $\gamma_c$ decoherence the excited state dependence was recently discussed in works \cite{Zhang:2007,Wenguo:2010} dealing with the time-dependant part of the Raman-Ramsey fringes reported earlier in \cite{Hemmer:1989,Shahriar:1997}.\\
\indent A perturbative expansion of Eq.~(\ref{Pulsed-Dark-Resonance}) in the Raman detuning parameter is required to derive the correct non linear behavior of the frequency-shift versus the $\Delta_0$ detuning, as reported in Eq.~(\ref{pulsed-fringe-frequency-shift}) of Appendix B. Those dependencies (solid dots) are plotted in Fig.~\ref{expanded-fringe-shift} for $\Omega_{2}/\Omega_{1}\sim0.75$ and compared to the numerical tracking shift of the fringe minimum from Eq.~(\ref{Pulsed-Dark-Resonance}) (solid lines). A non linear behaviour with small oscillations around the optical resonance is correctly described. Increasing the  $T$ time, there is a small rotation of the frequency-shift around the $\Delta_{0}=0$ common mode detuning as shown in Fig.~\ref{expanded-fringe-shift}(b). This dependence appears only for a non vanishing  $\gamma_{c}$ decoherence rate. For a given value of $T$, at specific values of the optical detuning, an exact cancelation of the frequency-shift takes place as seen in Fig.~\ref{expanded-fringe-shift}(b).

\section{Conclusion}
Three-level systems interacting with two coherent laser fields give rise to many phenomena, such as Autler-Townes doublet, Dark/EIT Resonance, and Fano-Feshbach transition. In all of them the Raman coherence, playing an essential role, is highly sensitive to the parameters of the laser interaction and of the decoherence processes. The present work explored how those parameters may be tuned for future optical clock devices providing sensitive measurements of energy levels shifts in neutral atoms clocks based on either fermionic or bosonic atomic species for example in dipolar traps \cite{Derevianko:2011} or in trapped ion clocks \cite{Leibfried:2003}.\\
\indent The three-level phenomena are well described by the formalism of the Bloch's equations in a semi-classical density matrix representation. From the exact resolution of Bloch's equations we have derived general analytic expressions of the resonance lineshapes observed in the steady-state of different atomic observables. We have examined the linewidth and power
broadening of the two-photon resonance. The precise dependence of the frequency shift associated to the Fano-Feshbach
transition or to the EIT resonance, not appearing in a
perturbation treatment, was discussed.
\indent We have examined the resonance fluorescence and frequency-shift for a pulsed laser configuration in the adiabatic regime. The analytical but asymptotic solution allows us to write the lineshape solution of the narrow quantum resonance leading to formation of Dark Resonance fringes.
The pulsed sequence overcomes the power broadening mechanism of the continuous-wave excitation while allowing high contrasted signals in a saturation regime. In the weak field limit, signals in
pulsed regimes are two times the continuous-wave signals, except for the decrease due to the Raman coherence relaxation within the atomic free evolution.
The lower limit to the resonance linewidth is $1/2$T to be compared to the $\Gamma_{eff}$ limit in the steady-state case. The resonance shift is still proportional to the $\gamma_c$ decoherence as for the steady state regime, but it is now diluted over the Ramsey time.\\
\indent An important and original result of our analysis is that different atomic or molecular quantum observables (excited population, clock-state populations and Raman coherence) experience different non-linear lineshapes and therefore different shifts of the clock resonance.  Depending on the atomic parameters, the cw shifts of those observables may greatly differ in slopes, magnitude or lineshape. The shift amplitudes are also strongly related to the values of the decay channels, either balanced or unbalanced.   For instance, while a large clock-population inversion is produced for unbalanced decay channels,  a dispersive lineshape is associated to its frequency shift. The detection of the Raman coherence exhibits a larger contrast, but it suffers usually of a systematic shift larger by an order of magnitude than that associated to the population detection. In the case of a very small decay rate for the Raman coherence,  destructive interferences strongly reduce the cw frequency-shift sensitivity of the excited state compared to others observables. Thus the choice of the observable is very important for the proper operation of a three-level atomic or molecular clock.  Additional technical constraints are associated to the requested detection tools of the chosen observable. For instance tracking the excited state population fraction by monitoring  spontaneous emission or light transmission represents a sort of non-demolition quantum measurement avoiding the destructive readout associated to the lower-state clock projection.\\
\indent The present solution can be extended to the case of a train of laser pulses having different frequencies,
optical detunings and phase steps as in refs \cite{Dalton:1986,Vandersypen:2005}, in order to
design more elaborated combinations of optical transient nutations and free evolutions and to explore  more
efficient detection scheme for very narrow transitions. For a treatment of the atom or molecule motion in matter wave interferometry~\cite{Berman:1997}, for instance based on a pulsed EIT/Raman interaction with a resonant excitation scheme and using stimulated photon recoils to create a beam splitter, the Bloch's set of Eq.~(\ref{set-Bloch}) requires both the introduction of the recoil shift into the laser detunings and the recoil spread among different atomic momentum classes produced by spontaneous emission yielding to a full set of quantum coupled equations~\cite{Aspect:1989}.

\section*{Acknowledgement}

We would like to deeply thank Christof Janssen for a careful reading of the paper. T.Z.W is also extremely grateful to LPMAA for making possible the final completion of this work. T.Z.W would like to address his best acknowledgments to J. Ye from JILA-NIST supporting over a long time our ideas on three-level coherence for optical clocks.

\appendix
\section {Generalized multi-photon steady-state rate solutions}

\begin{figure}[!t]
\centering
\resizebox{9.0cm}{!}{\includegraphics[angle=-90]{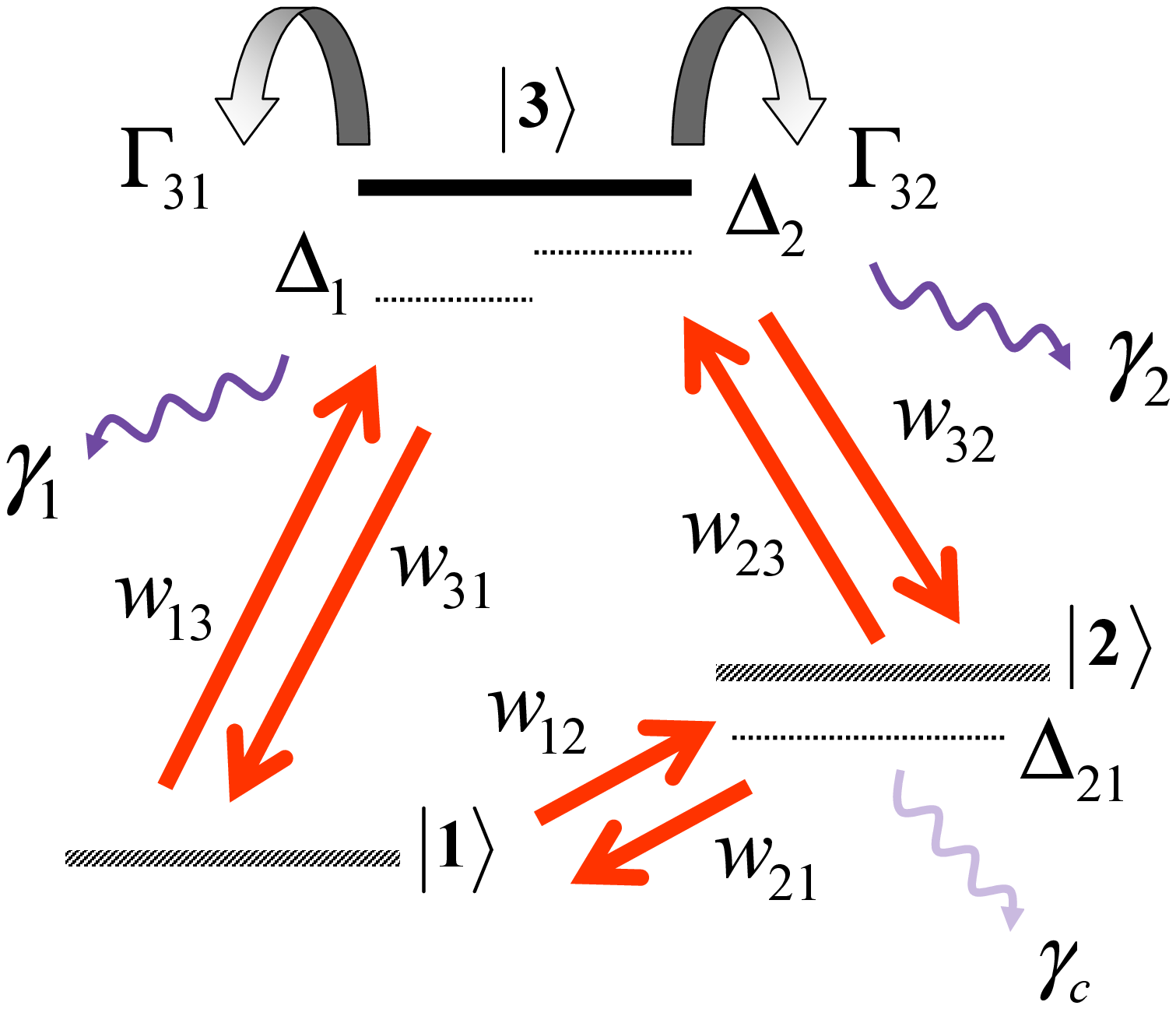}}
\caption{(Color online) Definition of the $w_{ij}$ transition rates within the three-level $\Lambda$
configuration}
\label{scheme-transition-rates}
\end{figure}
The lineshape expressions of Eq.~(\ref{fluorescence-solution}) and Eq.~(\ref{populations}) may be recast in a different mathematical form using
 the generalized multi-photon rate solution of ref.~\cite{Swain:1980}, later employed in ref.~\cite{Swain:1982} to establish conditions for the coherent population trapping in the steady-state regime. That analysis is based on three coupled equations for the $P_{ii}(Z)$, with $i=1,2,3$, occupation probabilities of the three levels, equations written in the Laplace space of variable $Z$. This approach shown in Fig.~(\ref{scheme-transition-rates}) allows a reduction in the size of the linearly-coupled equation system to be solved. The three level populations are obtained as following:
\begin{equation}
\rho_{ii}=lim_{t\rightarrow\infty}P_{ii}(t)=lim_{Z\rightarrow0}Z P_{ii}(Z).
\label{}
\end{equation}
From the steady-state solutions of refs.~\cite{Swain:1980,Swain:1982} we obtain
\begin{equation}
\begin{split}
\rho_{33}&=\frac{w_{12}w_{23}+w_{13}(w_{12}+w_{23})}{3w_{13}w_{12}+\Gamma_{32}W_{11}+\Gamma_{31}W_{22}+3w_{23}W_{33}},\\
\rho_{22}&=\frac{w_{12}w_{13}+\Gamma_{31}w_{12}+\Gamma_{32}(w_{13}+w_{12})+w_{23}W_{33}}{3w_{13}w_{12}+\Gamma_{32}W_{11}+\Gamma_{31}W_{22}+3w_{23}W_{33}},\\
\rho_{11}&=\frac{w_{12}w_{13}+\Gamma_{32}w_{12}+\Gamma_{31}(w_{23}+w_{12})+w_{23}W_{33}}{3w_{13}w_{12}+\Gamma_{32}W_{11}+\Gamma_{31}W_{22}+3w_{23}W_{33}}.
\end{split}
\label{steady-state-swain-solutions}
\end{equation}
where the $w_{ij}$ transition rates between each pair of states are defined in
Fig. 16 and where
\begin{equation}
\begin{split}
W_{11}&=w_{13}+2w_{12},\\
W_{22}&=w_{23}+2w_{12},\\
W_{33}&=w_{13}+w_{12}.
\end{split}
\end{equation}
The expressions for the  two-photon $(w_{12}\propto\Omega_{1}^{2}\Omega_{2}^{2})$ and  one-photon $(w_{13}\propto\Omega_{1}^{2},w_{23}\propto\Omega_{2}^{2})$ transitions rates  are
\begin{equation}
\begin{split}
w_{12}=&2Re\{\frac{\Omega_{1}^{2}\Omega_{2}^{2}}{\xi_{31}\xi_{21}\xi_{32}+\Omega_{1}^{2}\xi_{31}+\Omega_{2}^{2}\xi_{32}}\},\\
w_{13}=&2Re\{\frac{\Omega_{1}^{2}}{\xi_{31}+\frac{\Omega_{2}^{2}}{\xi_{21}+\frac{\Omega_{1}^{2}}{\xi_{32}}}}\}-w_{12},\\
w_{23}=&2Re\{\frac{\Omega_{2}^{2}}{\xi_{32}+\frac{\Omega_{1}^{2}}{\xi_{21}+\frac{\Omega_{2}^{2}}{\xi_{31}}}}\}-w_{12}.
\end{split}\label{}
\end{equation}
The detuning parameters $\xi_{ij}$ are complex numbers depending on the system parameters
\begin{equation}
\begin{split}
\xi_{31}&=i\Delta_{1}+\gamma_{1},\\
\xi_{32}&=-i\Delta_{2}+\gamma_{2},\\
\xi_{21}&=i\Delta_{21}+\gamma_{c}=i(\Delta_{1}-\Delta_{2})+\gamma_{c}.
\end{split}\label{}
\end{equation}
The above expressions of the one and two-photon rates point out the light-shift contributions to the  eigenfrequencies of the three-level $\Lambda$ system.

\section{Dark Resonance fringe frequency-shift}

\subsection{First order expression of $\delta\omega_{33}^{fr}$}

 Eq.~(\ref{Pulsed-Dark-Resonance}) can be recast in the following form:
\begin{eqnarray}
\begin{split}
\rho_{33}(T)=\alpha_{33}+B_{33}\cos\left(U\right),
\end{split}\label{}
\end{eqnarray}
where
\begin{eqnarray}
\begin{split}
B_{33}&=\mu|\rho_{12}|e^{-\gamma_{c}T},\\
\mu&=\alpha_{33}\beta_{33},\\
U&=\delta_{r} T-\Phi_{33}.
\end{split}\label{functions}
\end{eqnarray}
To establish the frequency-shift, a track of the extremum of Eq.~(\ref{Pulsed-Dark-Resonance}) produced by a differentiation versus $\delta_{r}$ leads to the following
expression:
\begin{equation}
\rho^{'}_{33}=\alpha_{33}^{'}+\sqrt{B_{33}^{'2}+B_{33}^{2}U^{'2}}sin(U+\theta_{33}),
\label{max}
\end{equation}
with
\begin{equation}
\left\lbrace
\begin{split}
\theta_{33}&=-\arctan(\dfrac{B_{33}^{'}}{B_{33}U^{'}}), \;\; if \;\; T-\Phi_{33}^{'}<0,\\
\theta_{33}&=-\arctan(\dfrac{B_{33}^{'}}{B_{33}U^{'}})+\pi,\;\; if \;\; T-\Phi_{33}^{'}>0.
\end{split}\right.\label{differentiation}
\end{equation}
The $\rho^{'}_{33}$ derivative vanishes at
\begin{eqnarray}
\begin{split}
U+\theta_{33}+\eta_{33}=0,
\end{split}\label{cancellation}
\end{eqnarray}
where
\begin{eqnarray}
\begin{split}
\eta_{33}=\arcsin(\dfrac{\alpha_{33}^{'}}{\sqrt{B_{33}^{'2}+B_{33}^{2}U^{'2}}}).
\end{split}\label{}
\end{eqnarray}
Rewriting the cancelation condition as:
\begin{eqnarray}
\delta_{r} T-\Phi_{33}+\theta_{33}+\eta_{33}=0,
\label{cancelation}
\end{eqnarray}
we apply a first order expansion of Eq.~(\ref{cancelation}) into the $\delta_{r}$ Raman detuning leading to the following $\delta\omega_{33}^{fr}$ frequency shift:
\begin{equation}
\delta\omega_{33}^{fr}\equiv\delta_{r}=\dfrac{\Phi_{33}(0)-\theta_{33}(0)-\eta_{33}(0)}{T-\Phi_{33}^{'}(0)+\theta_{33}^{'}(0)+\eta_{33}^{'}(0)}.
\end{equation}
If we neglect $\theta_{33}(0)$ and $\eta_{33}(0)$ terms, the discrepancy is respectively $+6\%$ and $-13\%$. If we neglect the $\theta_{33}^{'}(0)$ or the $\eta_{33}^{'}(0)$ term,
the error is below $+0.4\%$. Thus at the $1\%$ level accuracy and at a common mode detuning small compared to spontaneous emission rates, we can rewrite the frequency-shift expression as
\begin{equation}
\delta\omega_{33}^{fr}\approx\dfrac{\Phi_{33}(0)-\theta_{33}(0)-\eta_{33}(0)}{T-\Phi_{33}^{'}(0)}.
\label{pulsed-fringe-frequency-shift}
\end{equation}

\subsection{Derivative terms}

This subsection reports the functions appearing in Eq.~(\ref{pulsed-fringe-frequency-shift}) and required to determine  the frequency-shift $\delta\omega_{33}^{fr}$ as a function of the common mode detuning $\Delta_{0}$ and the system parameters.
The additional functions may be obtained from the first order derivatives of the $\mu,\Phi_{33},|\rho_{12}|,\alpha_{33}$ functions calculated at $\delta_{r}=0$. In order to simplify the mathematics, those functions are here reported for the  case of a pure radiative process $\gamma_{1}=\gamma_{2},\widetilde{\gamma}_{1}=\widetilde{\gamma}_{2}$, and supposing $\Delta_{1}=\Delta_{2}=\Delta_{0}$.
\begin{equation}
\begin{split}
\Phi_{33}(0)=-Arctan\left[\frac{\gamma_{c}\overline{\Delta}}{\gamma_{eff}}\right].
\end{split} \label{phi}
\end{equation}
\begin{eqnarray}
 \mu^{'}(0)=\mu(0)\frac{\Delta_0 \left( \tilde{\gamma_1} \Gamma  +3 \Omega_1^2-3 \Omega_2^2\right)}{\tilde{\gamma_1}\gamma_1\left( \tilde{\gamma_1}\Gamma  +3 \Omega_1^2+ 3\Omega_2^2\right)}.
 \label{muprimenew}
\end{eqnarray}
\begin{widetext}
\begin{eqnarray}
\begin{split}
\alpha_{33}^{'}(0)&=\frac{4 \Delta_0 \Omega_2^2\left(\Gamma  \tilde{\gamma_1}+\text{$\Delta $n}\left( \Gamma  \tilde{\gamma_1}+6\Omega_1^2 \right)\right)}{\Gamma  \tilde{\gamma_1} \left(\Gamma  \tilde{\gamma_1}+3 \left(\Omega_1^2+\Omega_2^2 \right)\right)^2} +\frac{\tilde{\gamma_1} \left(\Omega_2^2-\Omega_1^2\right)}{\left( \tilde{\gamma_1}{}^2\Gamma +3 \tilde{\gamma_1} \left( \Omega_1^2+\Omega_2^2\right)\right)\Gamma_{\text{eff}}^2} \times\\
& \left[ \Delta_{11} S^{11}+\Delta_{22} S^{22} -\gamma_c \overline{\Delta  } \overline{\gamma}_{22}  S^{22}+\gamma_c \Delta_0 \frac{2 \overline{\Omega}_1^2}{\Gamma } \left(\overline{\gamma}_{11} S^{11}+\overline{\gamma}_{22} S^{22}\right)+\text{$\Delta$n} \left(\Delta_f-\frac{8 \Delta_0 \overline{\Omega}_1^2 \left(\Gamma_{\text{eff}}^2-\gamma_c^2\right)}{4 \Delta_0^2+(1+2 S) \Gamma^2}\right)\right.\\
&+\frac{4 \Delta_0 S \overline{\Omega}_1^2}{\left( 1+2S+\frac{4\Delta_0^2}{\Gamma^2}\right)^2 \Gamma^2}\left(\gamma_{22}^2 \left(2+\frac{\tilde{\gamma_1} \Gamma_{32}}{\Omega_2^2}\right)-\gamma_{11}^2 \left(2+\frac{\tilde{\gamma_1} \Gamma_{31}}{\Omega_1^2}\right)\right)-\frac{2\Delta_0 \gamma_{22}^2 \Gamma_{32} S^{\Lambda }}{\Gamma \Omega_2^2} \\
&\left. +\frac{4 \Delta_0 \overline{\gamma}_{22} S^{22}\Gamma_{32} \left(2 \Delta_0 \overline{\Delta  }  \gamma_c+\Gamma
\gamma_{\text{eff}}\right) }{\Gamma_{32}\left(4 \Delta_0^2 + \Gamma^2 \right)+4\Gamma \Omega_2^2} \right]  .
\end{split}
\label{alphaprime}
\end{eqnarray}
\begin{eqnarray}
\Phi_{33}^{'}(0)=-\dfrac{\overline{\Delta}^2 \gamma_c \left(2 \gamma_1^2 \widetilde{\gamma_1}-2 \Delta_0^2 \gamma_c+\gamma_1 \widetilde{\gamma_1} \gamma_c\right)+\gamma_{eff} \left(2 \gamma_1 \widetilde{\gamma_1} (\gamma_1+ \overline{\Omega}_1^{2} \gamma_c)-2 \Delta_0^2 \gamma_{eff}+\gamma_1 \widetilde{\gamma_1} \gamma_{eff}\right)}{2 \gamma_1^2 \widetilde{\gamma_1} \left(\overline{\Delta}^2 \gamma_c^2+\gamma_{eff}^2\right)}.
\label{phiprime}
\end{eqnarray}
\begin{eqnarray}
\begin{split}
|\rho_{12}|^{'}(0)=|\rho_{12}|(0)\left( \frac{\Delta_f}{\Gamma_{\text{eff}}^2} -\frac{ 8 \Delta_0 \left(4 \Delta_0^{2} \gamma_c  S+\Gamma  \left(\gamma_{\text{eff}}^2+3 \gamma_c \gamma_{\text{eff}} S-\gamma_c^2
 (1+2 S)\right)\right)\overline{\Omega}_1^2}{\left( 1+2 S + 4\Delta_0^2 /\Gamma^2\right)^2 \Gamma^3 \Gamma_{\text{eff}}^2} \right.\\
  \left. +\frac{\bar{\Delta }  \left(\gamma_c -\gamma_{\text{eff}}+2 \overline{\Omega}_1^2 \gamma_c^2 /\Gamma \right) }{ \left(\bar{\Delta }^2 \gamma_c^2+\gamma_{\text{eff}}^2\right) } +\frac{8 \Delta_0 \overline{\Omega}_1^2}{ \Gamma^2 \left( 1+2 S + 4\Delta_0^2 /\Gamma^2\right)}\right).
\end{split}
\label{rho12prime}
\end{eqnarray}
\end{widetext}
These perturbative expressions were used to determine the $\delta\omega_{33}^{fr}$ frequency-shift plotted in Fig.~\ref{expanded-fringe-shift} when $\Delta_{0}\leq\Gamma$.


\begin{thebibliography}{1}
\bibitem{note0} All quantities with an overbar are dimensionless.
\end{thebibliography}
\end{document}